\documentclass{emulateapj}
\usepackage{graphicx}
\usepackage{natbib}
\pdfoutput=1
\bibpunct{(}{)}{;}{a}{}{,} %

\usepackage{amsmath}
\usepackage{amssymb}
\usepackage{wasysym}
\usepackage{url}

\begin{document}

\title{Maps of Evolving Cloud Structures in Luhman 16AB from HST Time-Resolved Spectroscopy.}

\author{Theodora Karalidi}
\affil{Steward Observatory, Department of Astronomy, University of Arizona, 933 N. Cherry Ave, Tucson, AZ 85721, USA, \\ \textit{tkaralidi@email.arizona.edu}}

\author{D\'aniel Apai}
\affil{Steward Observatory, Department of Astronomy, The University of Arizona, 933 N. Cherry Ave, Tucson, AZ 85721, USA \\
Lunar and Planetary Laboratory, University of Arizona,1629 E University Blvd, AZ 85721, USA \\ 
Earths in Other Solar Systems Team}

\author{Mark S. Marley}
\affil{NASA Ames Research Center, MS-245-3, Moffett Field, CA 94035, USA }

\author{Esther Buenzli}
\affil{Institute for Astronomy, ETH Z{\"u}rich, Wolfgang-Pauli-Str. 27, 8093 Z{\"u}rich, Switzerland}

\begin{abstract}

WISE J104915.57-531906.1 is the nearest brown dwarf binary to our Solar 
system, consisting of two brown dwarfs in the L/T transition: Luhman 16A \& B. 
In this paper we present the first map of Luhman 16A, and maps of Luhman 16B 
for two epochs. Our maps were created by applying \textit{Aeolus}, a Markov--Chain 
Monte Carlo code that maps the top--of--the--atmosphere (TOA) structure of brown dwarf and 
other ultracool atmospheres, to light curves of Luhman 16A \& B using the Hubble 
Space Telescope's G141 and G102 grisms. \textit{Aeolus} retrieved three or four spots 
in the top--of--the--atmosphere of Luhman 16A \& B, with a surface coverage of 
19\%--32\% (depending on an assumed rotational period of 5 hr or 8 hr) or 21\%--38.5\% (depending on the 
observational epoch) respectively.  The brightness temperature of the 
spots of the best--fit models was $\sim$200 K hotter than the 
background TOA. We compared our Luhman 16B map with the only previously 
published map. Interestingly, our map contained a large, cooler ($\Delta T \sim51$ K) than the 
background TOA spot that lay at low latitudes, in agreement with the previous Luhman 16B map. 
Finally, we report the detection of a feature 
reappearing in Luhman 16B light curves that are separated by tens of hundreds of rotations from each other. We 
speculate this feature is related to TOA structures of Luhman 16B.

\end{abstract}


\section{Introduction}

WISE J104915.57-531906.1, also known as Luhman 16AB \citep[][]{luhman13}, is the nearest 
brown dwarf binary to our Solar system, at a distance of $1.9980\pm0.0004$ pc 
\citep[][]{sahlmann15}.
Luhman 16AB is composed of two brown dwarfs in the L/T transition: Luhman 16A, an L$8\pm1$ [L7.5] dwarf 
and Luhman 16B a T$1.5\pm2$ [T0.5] dwarf (\citet[][]{kniazev13} [\citet[][]{burgasser13}]). 
Luhman 16B has a rotational period of 4.87$\pm$0.01 hr \citep[][]{gillon13}, or 
5.05$\pm$0.10 hr \citep[][]{burgasser14}, and an inclination $i<30^\circ$ \citep[][]{crossfield14}. 
\citet[][]{buenzli15b} using HST observations, reported a rotational period between 4.5 and 5.5 hr for 
Luhman 16A, while ground-based, spatially resolved observations 
by \citet[][]{mancini15} suggested a longer, $\sim$ 8 hr rotational 
period. The latter, is longer than the maximum rotational period suggested by $v\sin i$ 
observations by \citet[][]{crossfield14}, assuming a $\sim$1$R_{Jup}$ radius \citep[see,][]{buenzli15b}. 
Luhman 16A has a yet undefined inclination.  
 \citet[][]{faherty14} using bolometric luminosities found that the two components 
have a similar T$_{eff}$ (Luhman 16A: 1310$\pm$30K and Luhman 16B: 1280$\pm$75K), consistent 
with previous observations by \citet[][]{kniazev13}.

\citet[][]{gillon13} performed partially resolved observations of Luhman 16AB over multiple rotations and 
reported strong night--to--night variations of the observed 
light curves and a large peak--to--peak amplitude ($\sim$10\%). Resolved observations 
showed that Luhman 16B is responsible for the observed variability \citep[][]{biller13,burgasser14,buenzli15}, 
while Luhman 16A did not appear variable above $\sim$0.4\% \citep[see, e.g., ][]{buenzli15}. 
Resolved observations by \citet[][]{burgasser13} showed that both components are 
red and underluminous in the J--band, supporting the idea that clouds are responsible for the 
observed light curve variability. Later observations by 
\citet[][]{buenzli15b} using the Hubble Space Telescope (HST)/WFC3 G102 grism reported a 
significant variability for both Luhman 16A (peak--to--peak amplitude of $\sim$4.5\%) and 16B 
(peak--to--peak amplitude of $\sim$9.3\%), cautioning about the implications this has to the unambiguous 
characterization of the two components in unresolved observations.

\citet[][]{radigan14}  and \citet[][]{radigan14b} found that 39$^{+16}_{-14}$\% of 
brown dwarfs in the L/T transition are variable, and showed that high--amplitude variability ($\gtrsim$ 2\%) is 
uncommon in the L/T transition (24$^{+11}_{-9}$\%). 
\citet[][]{metchev14} reached a similar conclusion showing that 
few L/T dwarfs have amplitudes larger than 1\%--2\%. The fact that both Luhman 16A and B 
present high--amplitude variability makes this brown dwarf binary a unique object.  
\citet[][]{buenzli15b} performed a statistical analysis and found that 
the combined probability for Luhman 16A and B to be both variable is 
$\lesssim10$\%. The reason why both components present a large amplitude variability 
is yet unclear.

\citet[][]{crossfield14} used Doppler imaging to produce the first global map of Luhman 16B.  
This first map indicated a complex cloud structure, with darker and brighter areas across the  
globe, which \citet[][]{crossfield14} interpreted as observations of thinner and thicker clouds that 
allowed the observations to penetrate deeper or shallower in the atmosphere. This is in 
agreement with the general picture suggested by e.g., \citet[][]{apai13}  for other early T--dwarfs.

In \citet[][]{karalidi15} we presented \textit{Aeolus}, a Markov Chain Monte Carlo (MCMC) code 
that can map the top--of--the--atmosphere (hereafter, TOA) structure of an ultracool atmosphere, 
per observational wavelength, using observed rotational light curves. In Section~\ref{sect:Aeolus} 
we present a brief  description of \textit{Aeolus}. For a detailed description of \textit{Aeolus}, and 
its validation, we refer the reader to \citet[][]{karalidi15}. 

In this paper we used the Luhman 16B light curves of \citet[][]{buenzli15, buenzli15b} to map 
Luhman 16B's TOA structure integrating over the J-- and H--bands (cut at 1.66 $\mu$m) 
of the G141 grism and in the complete G102 grism. Using multi-wavelength observations 
to study an ultracool atmosphere is a powerful technique that allows us to 
probe different layers in the observed atmosphere. 
Contribution functions of early T--dwarfs suggest that the J-- and H--band, and the G102 grism 
probe very similar pressure levels (see Section~\ref{sect:luhm16b_maps}). The maps 
we made using these bands thus, provide us with hints of the evolution of the atmospheric 
structure of Luhman 16B between the two epochs of observations. 

Finally, we used the Luhman 16A curves of \citet[][]{buenzli15b} to create the first map 
of the TOA structure of Luhamn 16A. We created two maps for Luhman 16A, using the two 
proposed rotational rates by \citet[][]{buenzli15b} and \citet[][]{mancini15}. Using our 
adapted version of \textit{Aeolus} (see Sect.~\ref{sect:Aeolus}) we constrained the inclination 
of Luhman 16A for the two rotational periods.

This paper is organized as follows. In Sect.~\ref{sect:data} we present the 
dataset we use in this paper. In Sect.~\ref{sect:Aeolus} we introduce, briefly, \textit{Aeolus}. 
In Sect.~\ref{sect:single_spots} we present the Principal component analysis results of 
\citet[][]{buenzli15, buenzli15b}. In Sects.~\ref{sect:luhm16b_maps} and~\ref{sect:luhm16a_maps} 
we present \textit{Aeolus} maps of Luhman 16B and 16A respectively. Finally, in Sect.~\ref{sect:discussion} 
we present a discussion of our results and in Sect.~\ref{sect:concl} we present our conclusions.

\section{Data} \label{sect:data}

Luhman 16AB was observed on 2013 November 8 with HST/WFC3 with the G141 grism 
(from 1.1 to 1.66 $\mu$m)
by \citet[][]{buenzli15}, and on 2014 November 23 with HST/WFC3 with the G102 grism 
(from 0.8 to 1.15 $\mu$m) by 
\citet[][]{buenzli15b}. For a detailed description of the observations and data reduction we refer 
the reader to \citet[][]{buenzli15,buenzli15b}. Each set of observations covered approximately 1.5  
rotations of Luhman 16B. In Table~\ref{table:obs} we summarize the observations we used in this paper.

\begin{figure}
\centering
\includegraphics[height=60mm]{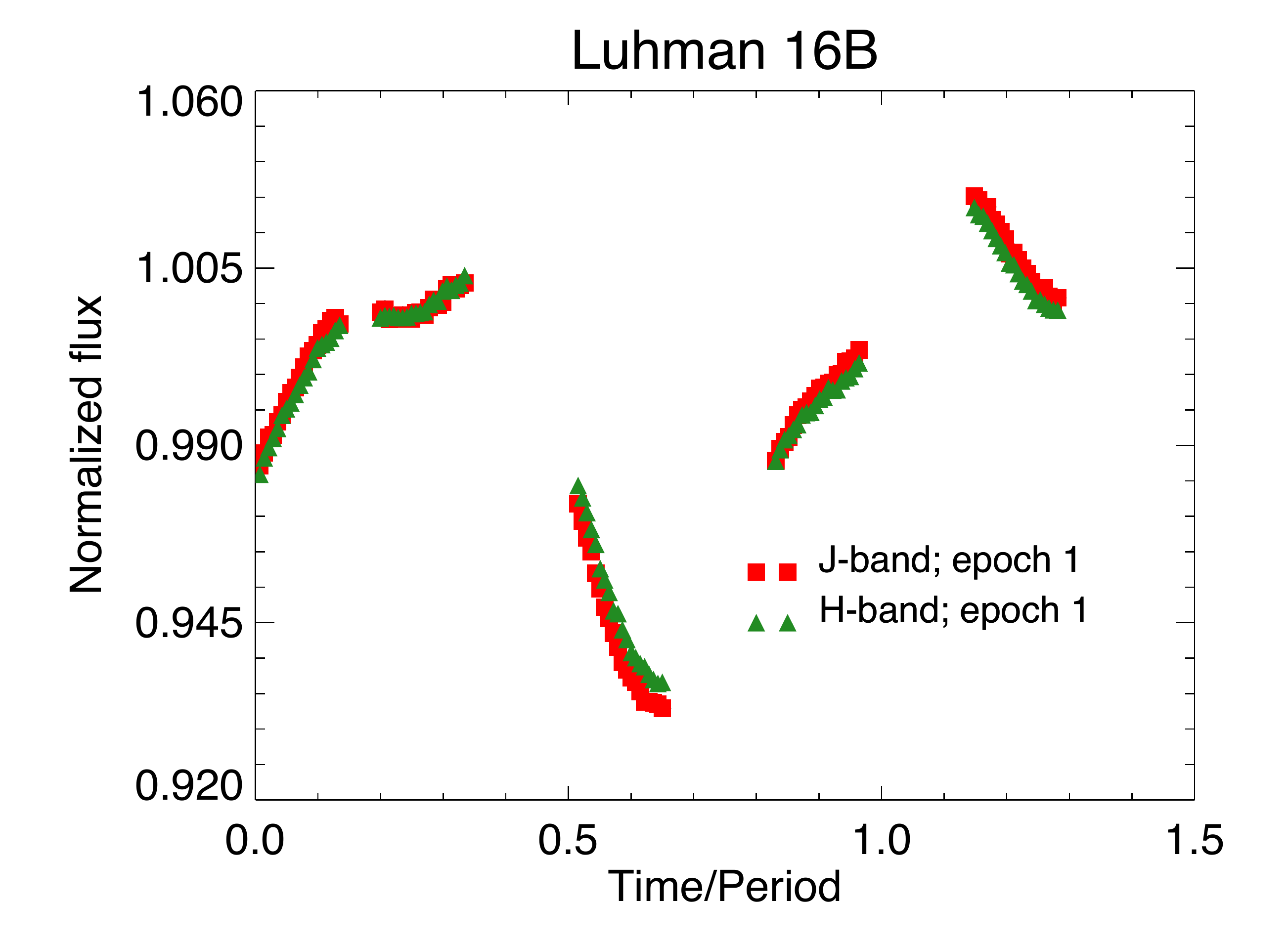}
\centering
\includegraphics[height=60mm]{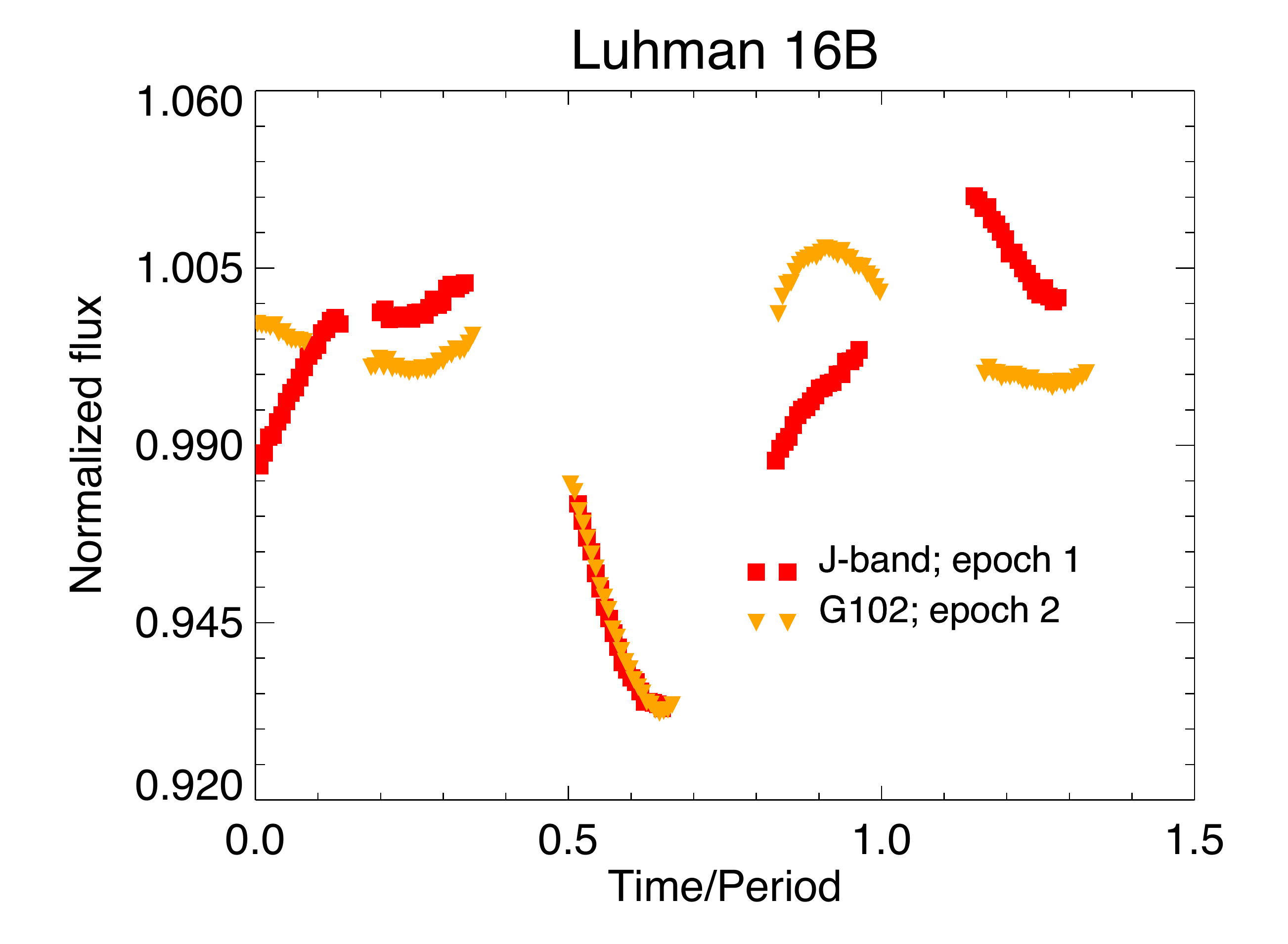}
\caption{Luhman 16B light curves from \citet[][]{buenzli15} and \citet[][]{buenzli15b}, assuming 
a rotational period of 5.05 hr \citep[][]{burgasser14}. 
Top panel: 2013 November 8, HST/WFC3  light curves in the J-- (red blocks) and H-- (green triangles) 
bands. Bottom panel: 2014 November 23, HST/WFC3, G102 light curves 
over the complete G102 grism (orange arrows). The J--band (red blocks) light curve is also plotted 
for comparison, shifted so that the trough around a rotational phase of 0.6 match.}
\label{fig:luhmanB_lc}
\end{figure}

Fig.~\ref{fig:luhmanB_lc} shows the Luhman 16 B light curves derived by integrating counts of \citet[][]{buenzli15} 
in the J-- (red blocks) and H-- (green triangles) bands (top panel), and 
of \citet[][]{buenzli15b} over the complete G102 grism (orange triangles, bottom panel). We note  
that Luhman 16B has a very rapidly evolving atmosphere, and we can detect evolution in the light curves 
from the first to the second rotation on the 2013 observations, and from the 2013, J-- and H--band 
light curves to the 2014 G102 light curve. 

\begin{figure}
\centering
\includegraphics[height=60mm]{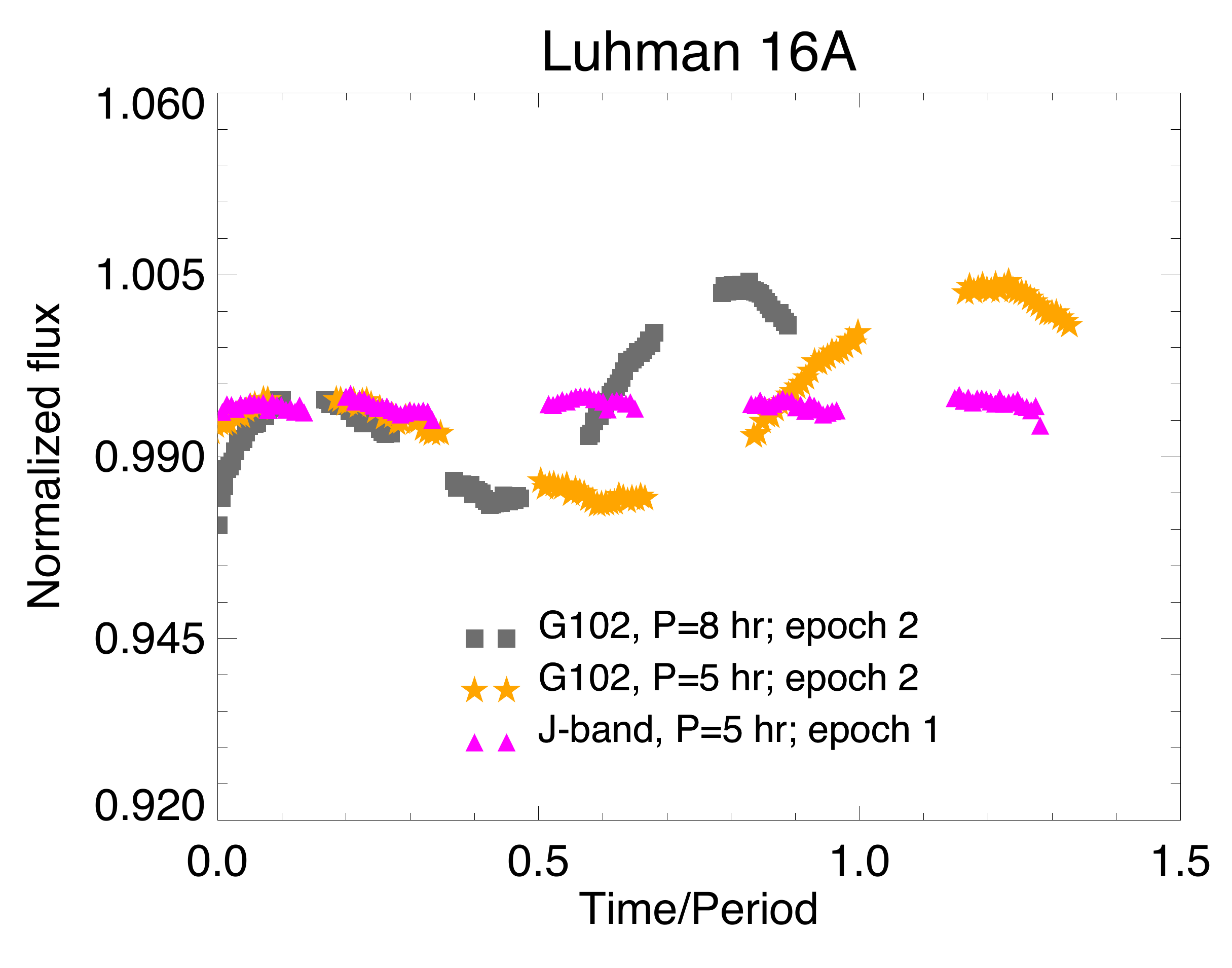}
\caption{Luhman 16A HST/WFC3, G102 light curve from \citet[][]{buenzli15b} (epoch 2) 
assuming a rotational period of 8 hr (grey blocks) and 5 hr (orange stars). 
The J--band (magenta triangles) light curve 
from \citet[][]{buenzli15} (epoch 1)  is over--plotted for comparison. }
\label{fig:luhmanA_lc}
\end{figure}

Finally, Fig.~\ref{fig:luhmanA_lc} shows the Luhman 16A 
curves derived by integrating counts over the complete G102 grism \citep[][]{buenzli15b}, 
assuming  a rotational period of 8 hr (grey blocks) and 5 hr (orange stars). The J--band 
(magenta triangles) light curve from \citet[][]{buenzli15} is also plotted for comparison.

\begin{table}[t]
\caption{Summary of observations used in this paper.}
\centering
\resizebox{.5\textwidth}{!} {
\begin{tabular}{c c c c c}
\hline
\hline
Original Paper & Epoch & Date of observations & Band \\
\hline
\citet[][]{buenzli15} & 1 & 2013 Nov. 8 & J \\
\citet[][]{buenzli15} & 1& 2013 Nov. 8 & H  \\
\citet[][]{buenzli15b} & 2 & 2014 Nov. 23 & G102 \\ 
\hline
\end{tabular}
}
\label{table:obs}
\end{table}


\section{Aeolus} \label{sect:Aeolus}
We mapped Luhman 16B using \textit{Aeolus}.  \textit{Aeolus} is an 
MCMC code that combines a Gibbs sampler with a random--walk Metropolis--within--Gibbs algorithm
to map the TOA (per observational wavelength) 
structure of ultracool atmospheres \citep[][]{karalidi15}. \textit{Aeolus} assumes 
that variations in the observed light curves are due to elliptical, spot--like 
features \citep[see,][]{karalidi15} and fits the number of spots as a free parameter. 
For every spot, \textit{Aeolus} fits the position (longitude and latitude), angular size, and 
contrast ratio to the background TOA. 

In this paper we allowed the contrast ratio of every spot to the background to vary 
between 0.01 and 2.0, and set the maximum allowed number of spots to 7. 
We varied \textit{Aeolus} in comparison to \citep[][]{karalidi15} to fit 
the inclination and limb darkening of our target as extra free parameters. 
As discussed in \citet[][]{karalidi15}, since there 
is no intrinsic reason why \textit{Aeolus} should prefer specific 
values of the parameters it fits (longitude, latitude, size and contrast ratio) 
over others, we assign a uniform prior ($p(\emph{x}) \sim1$) over 
their respective parameter ranges. We assume that the observational errors are 
nearly Gaussian, with known variances, and adopt a normal likelihood distribution 
($p(\emph{d}|\emph{x})\sim$exp[-$\chi^2$(\emph{x})/2]).

For every light curve we ran eight chains of 5,000,000 steps each. 
Removal of possible biases rising from our selection of initial conditions was done by 
removing the first 500,000 steps (10\%) of the chains \citep[see][and references therein]{karalidi15}. 
The choice of the best fitting model finally took into account the minimization 
of the Bayesian Information Criterion (BIC) \citep[][]{schwarz78}. When comparing two 
models the one with the smaller BIC is preferred. If the BIC of the two models is the 
same, the model with the fewer free parameters is preferred. 
As in \citet[][]{karalidi15}, to control that the solution on which our MCMC chains converged did not depend 
on our initial guesses, we used different initial condition for each chain and used the Gelman \& Rubin 
$\hat{R}$ criterion to control the convergence of the chains \citep[][]{gelmanrubin92}. 
To accept a solution we checked that $\hat{R}$ is always less than 1.2. 
Finally, we kept our sample size $N$ larger than 
the number of our fitted parameters $\kappa$ ($200\gtrsim N/\kappa\gtrsim 40$).
For a detailed description of \textit{Aeolus} we refer the reader to \citet[][]{karalidi15}.

\section{Single Spot--Component model} \label{sect:single_spots}

\citet[][]{buenzli15, buenzli15b} performed a Principal Component Analysis (PCA) on their 
data to determine the minimum number of independent spectral components necessary to 
reproduce their observed spectra. They concluded that for both sets of observations the observed 
spectral variability could be mostly characterized by a single component, implying that 
only two significant discrete photospheric structures (a ``background'' and a ``heterogeneity'') 
defined the TOA structure of these brown dwarfs. 
A similar conclusion was reached for 2MASSJ21392676+0220226 
and 2MASSJ0136565+093347 by \citet[][]{apai13}. 

We assumed that the ``heterogeneity'' component of the TOA is structured in 
the shape of elliptical, spot--like features (see Sect.~\ref{sect:Aeolus}).
We then assumed that the fact that all possible spots are dominated by one spectral component implies 
that their temperature difference to the background TOA was similar. 
Assuming that any brightness variations mapped by \textit{Aeolus} are due to the 
different temperatures of the areas observed, this implied that the contrast ratios of all the spots 
retrieved by \textit{Aeolus} should be similar. We thus adapted \textit{Aeolus} to keep the contrast 
ratio of all (possible) spots the same.

\section{Luhman 16B}\label{sect:luhm16b_maps}

We initially applied \textit{Aeolus} on the Luhman 16B light curves of Fig.~\ref{fig:luhmanB_lc}. 
These light curves cover approximately 1.3 to 1.4 rotations of Luhman 16B. The atmosphere 
of Luhman 16B appears to be highly active, causing the light curves to evolve already within 
one complete rotation (see, e.g., the change of the light curve shape between a rotational 
phase of 0.1 to 0.3  and 1.1 to 1.3 on the top panel of Fig.~\ref{fig:luhmanB_lc}). 
For this reason we split the light curves and fitted 
only the first complete rotation (rotational phase 0. to 1.).

Here we present the maps \textit{Aeolus} derived for the two epochs.
We followed \citet[][]{karalidi15}, and assumed that brightness variations across the TOA are due to 
the different temperature of the areas observed. We thus used the retrieved contrast ratios to calculate 
brightness temperature variations across the TOA. 

\textit{Aeolus} retrieved an inclination of $26^\circ\pm8^\circ$ for Luhman 16B, in agreement 
with \citet[][]{crossfield14} observations that suggested an inclination $i<30^\circ$.

\subsection{First epoch maps}

\begin{figure}
\centering
\includegraphics[height=68mm]{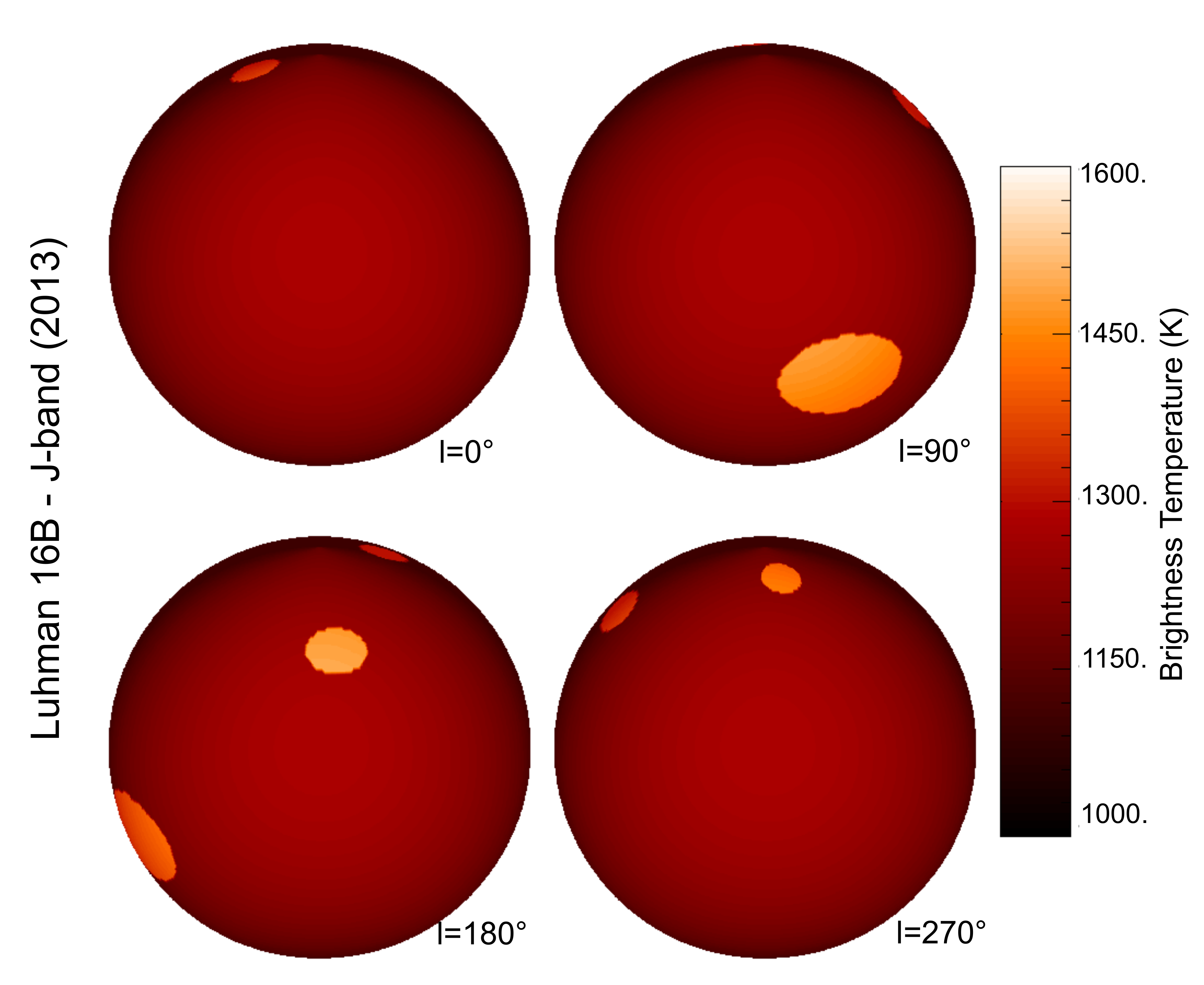}
\vspace{2pt}
\centering
\includegraphics[height=68mm]{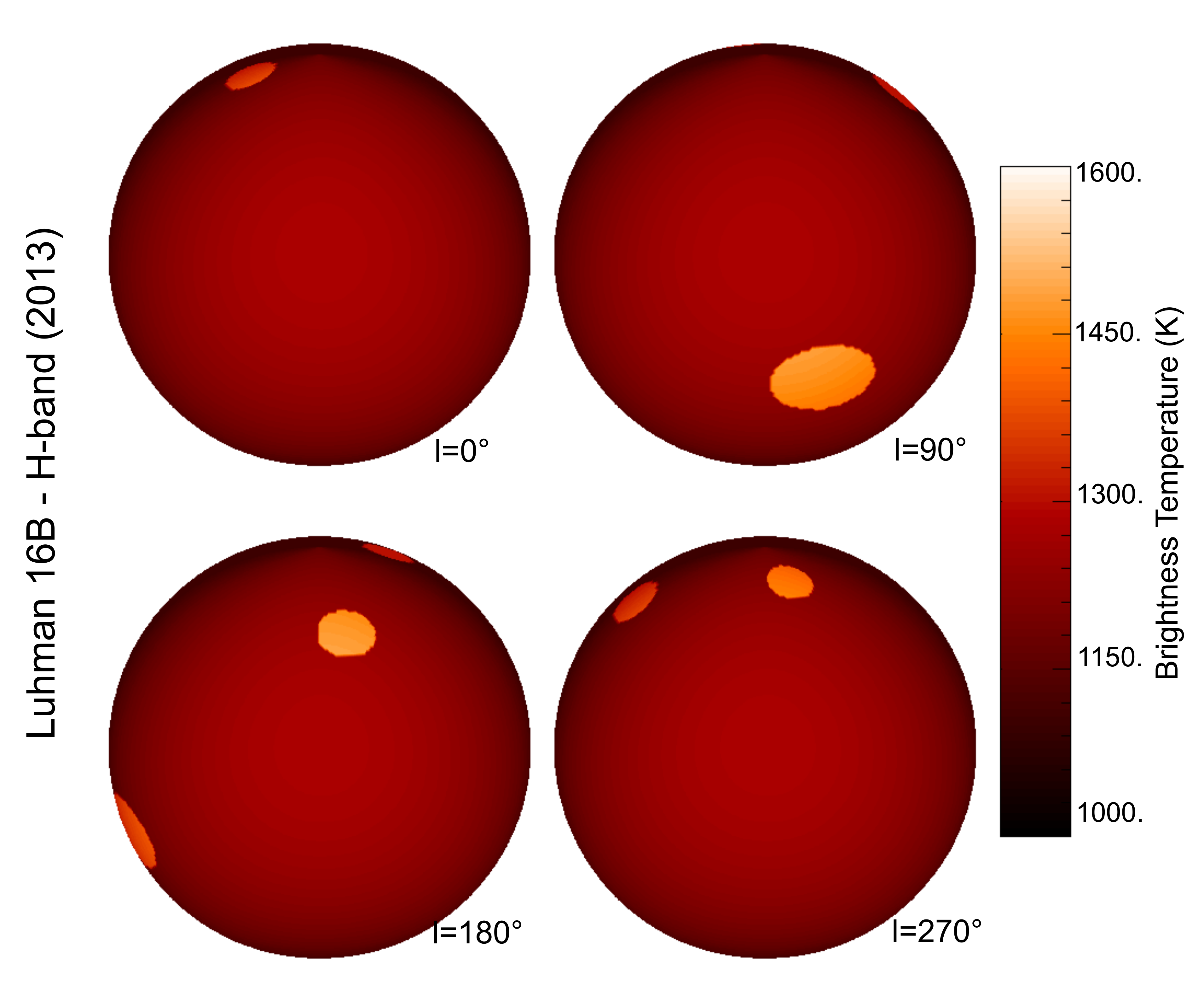}
\caption{Luhman 16B brightness temperature maps from applying 
\textit{Aeolus} on the J-- (top four panels) and the H--band (bottom four panels) 
light curves of Fig.~\ref{fig:luhmanB_lc}. The maps are centered 
at $0^\circ$ of longitude (upper left map), $90^\circ$ of longitude (upper right map), 
$180^\circ$ of longitude (lower left map) and $270^\circ$ of longitude (lower right map). }
\label{fig:luhmanB_jmap}
\end{figure}

Fig.~\ref{fig:luhmanB_jmap} (top four panels) shows the brightness temperature map of Luhman 16B in the J--band. 
\textit{Aeolus} retrieved three spots (BIC$\sim$25.9 vs 41.4 for two spots) with (longitude, latitude) = 
($113.47^\circ\pm6.46^\circ$, $31^\circ\pm6^\circ$), ($186.08^\circ\pm3.65^\circ$,  
$45^\circ\pm12^\circ$) and ($283^\circ\pm13^\circ$, $72^\circ\pm10^\circ$) and respective 
sizes of $38.89^\circ\pm0.80^\circ$, $24.45^\circ\pm1.46^\circ$ and $35.54^\circ\pm1.67^\circ$.
Assuming a background TOA temperature of 1280 K  \citep[][]{faherty14}, the retrieved spots had a 
temperature difference to the background TOA $\Delta T \sim 231\pm16$ K. 

\begin{figure}
\centering
\includegraphics[height=58mm]{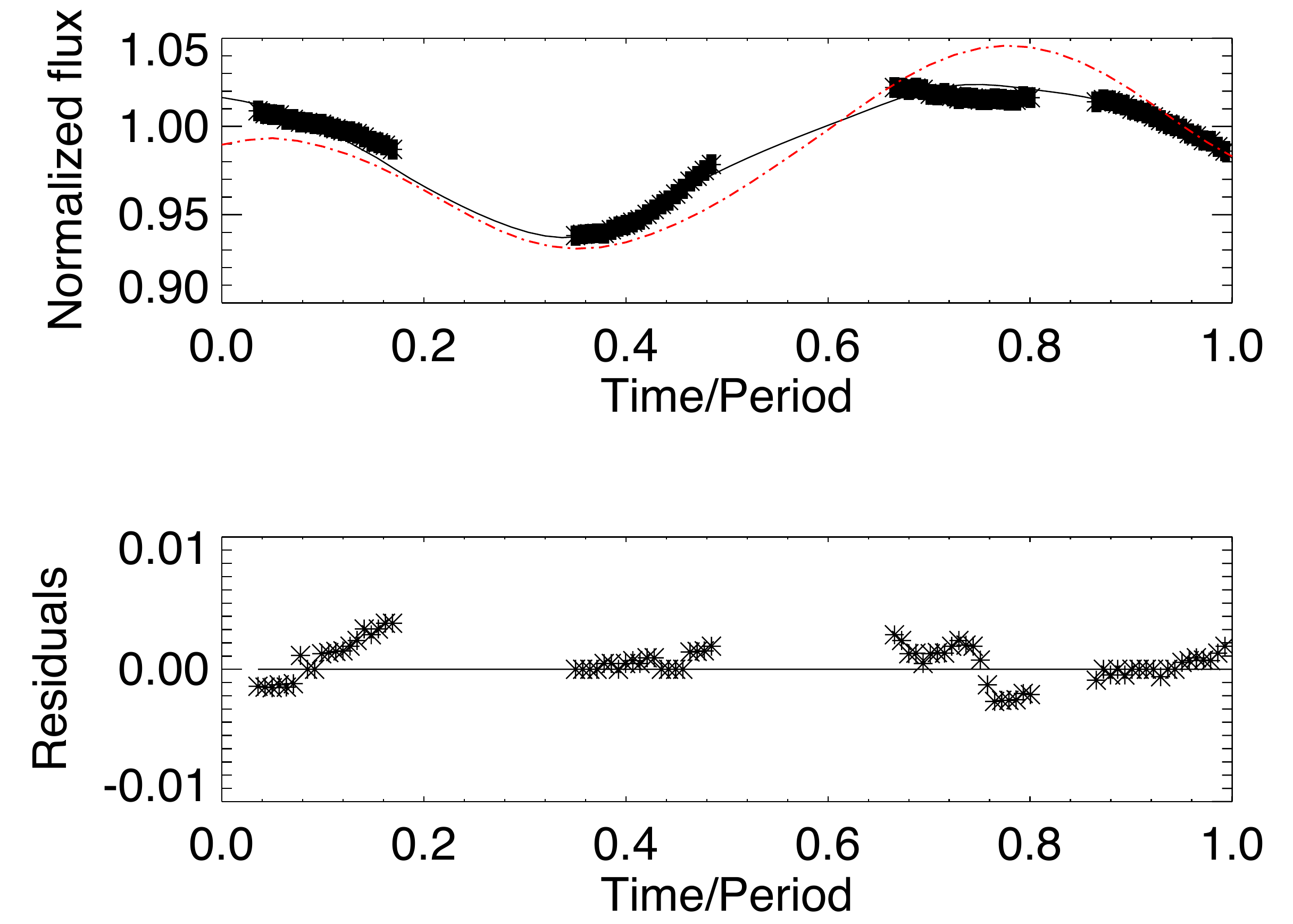}
\caption{Top panel: Normalized J--band light curve of Luhman 16B (stars) with 
error bars and best fit \textit{Aeolus} curve (black solid line). 
Note that the error bars are smaller than the symbols. Also plotted is the 
best fit 2-spot model (red dashed--dotted line) for comparison. 
Bottom panel: corresponding residuals. }
\label{fig:lon_histogram}
\end{figure}

\begin{figure}
\centering
\includegraphics[height=57mm]{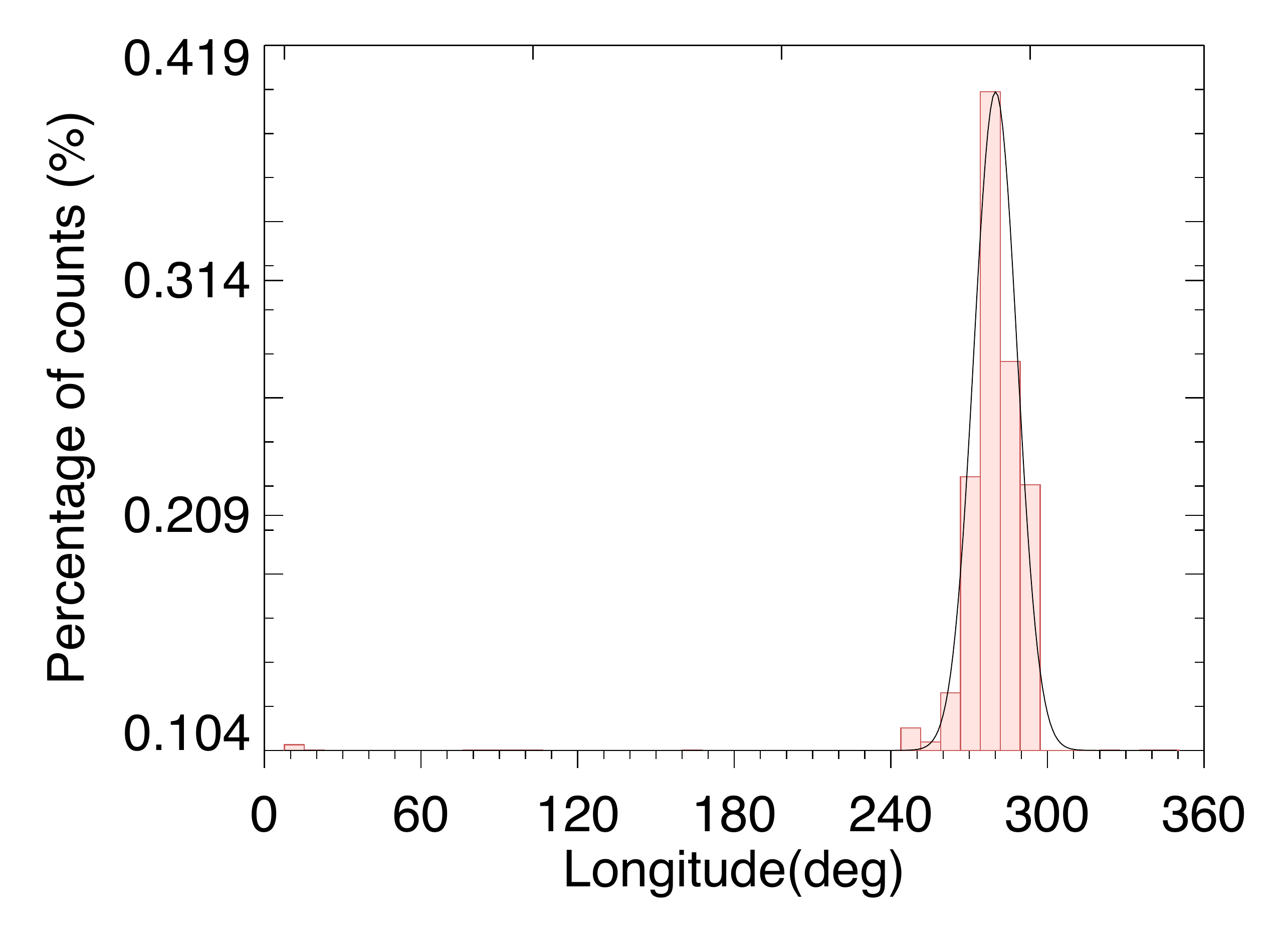}
\vspace{-2mm}
\centering
\includegraphics[height=55mm]{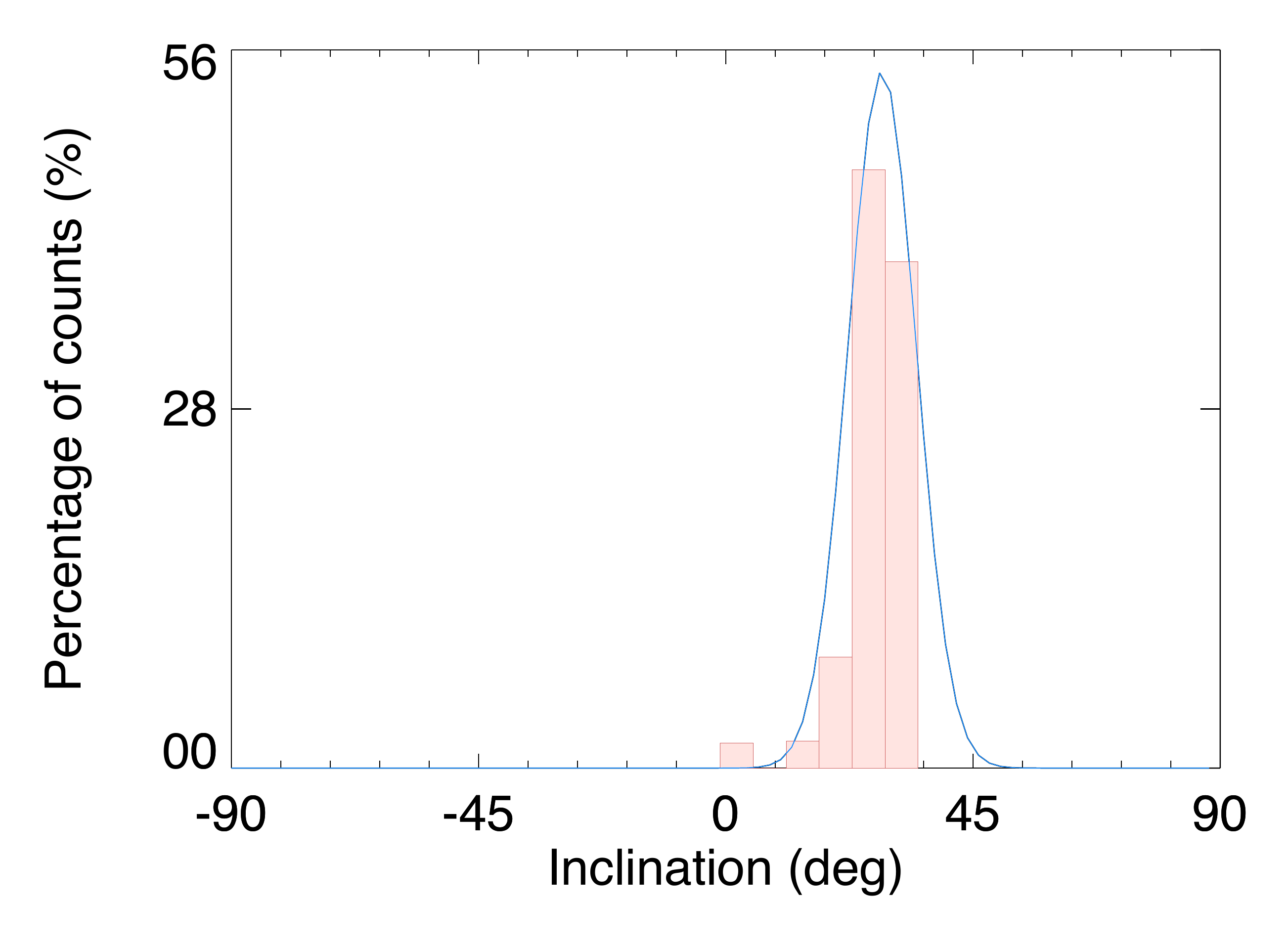}
\vspace{-2mm}
\centering
\includegraphics[height=55mm]{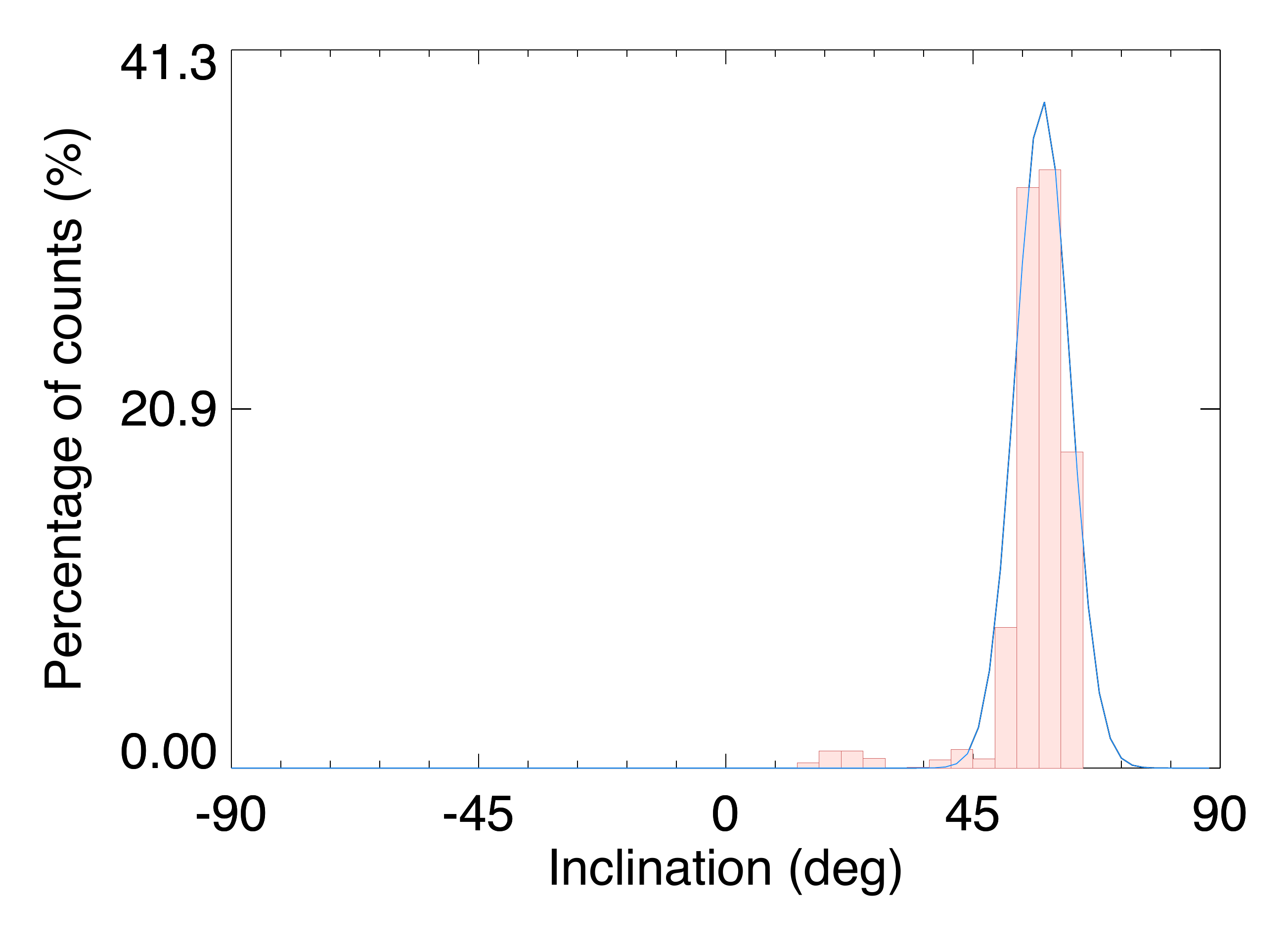}
\caption{
Top panel: Sample posterior distribution for the longitude of spot 
3 of Luhman 16B (J--band). Middle panel: Sample posterior distribution for the inclination 
of Luhman 16B (J--band). Bottom panel: Sample posterior distribution for the inclination of Luhman 16A (P= 5 hr). }
\label{fig:lon_histogramB}
\end{figure}

In Fig.~\ref{fig:lon_histogram} (top panel) we show the normalized J--band light curve (stars) 
with error bars and the best-fit \textit{Aeolus} curve (black line). In Fig.~\ref{fig:lon_histogram} (bottom panel) we 
show the corresponding residuals. The best-fit 2--spots \textit{Aeolus} curve (red dashed--dotted line) is also shown 
for comparison. In Fig.~\ref{fig:lon_histogramB} we show sample posterior distributions for the longitude of spot 3, the inclination of Luhman 16B (based on the J--band light curve) and the inclination of Luhman 16A assuming a 
rotational period of 5 hr.

To control the robustness of our results, and to take advantage of the complete light curve, 
we applied again \textit{Aeolus} on the J--band light curve 
between a rotational phase of 0.2 to $\sim$1.2. For a direct comparison with the previous fit, 
we present the longitudes shifted by $\sim70^\circ$ (rotational phase of 0.2).
\textit{Aeolus} retrieved three spots (BIC$\sim$60 vs 71 for two spots) with (longitude, latitude) = 
($192^\circ\pm12^\circ$, $30^\circ\pm15^\circ$), ($287^\circ\pm12^\circ$, $60^\circ\pm12^\circ$) and 
($376^\circ\pm8^\circ$, $-10^\circ\pm13^\circ$), and respective 
sizes of $26^\circ\pm1^\circ$, $37^\circ\pm1^\circ$ and $15.5^\circ\pm1.4^\circ$. 
The properties of the first two spots were in agreement with two spots of the previous fit. The third spot
 \textit{Aeolus} retrieved did not match the properties of the ``first spot'' of the previous fit (with 
 longitude of $113.47^\circ\pm6.46^\circ$), hinting to the rapid evolution of the TOA structure of 
 Luhman 16B.


Fig.~\ref{fig:luhmanB_jmap} (bottom four panels) shows the brightness temperature 
map of Luhman 16B in the H--band. 
\textit{Aeolus} retrieved three spots (BIC$\sim$24 vs 42 for four spots) with (longitude, latitude) = 
($107.27^\circ\pm5.48^\circ$, $28^\circ\pm8^\circ$), ($190.91^\circ\pm4.83^\circ$,  
$50^\circ\pm20^\circ$) and ($288.6^\circ\pm9.4^\circ$, $74^\circ\pm6^\circ$) and respective 
sizes of $32.42^\circ\pm0.80^\circ$, $25.02^\circ\pm1.13^\circ$ and $36.30^\circ\pm1.18^\circ$.
Assuming a background TOA temperature of 1280 K, the retrieved spots had a 
$\Delta T \sim 211\pm16$ K to the background TOA. 

\begin{figure}
\centering
\includegraphics[height=68mm]{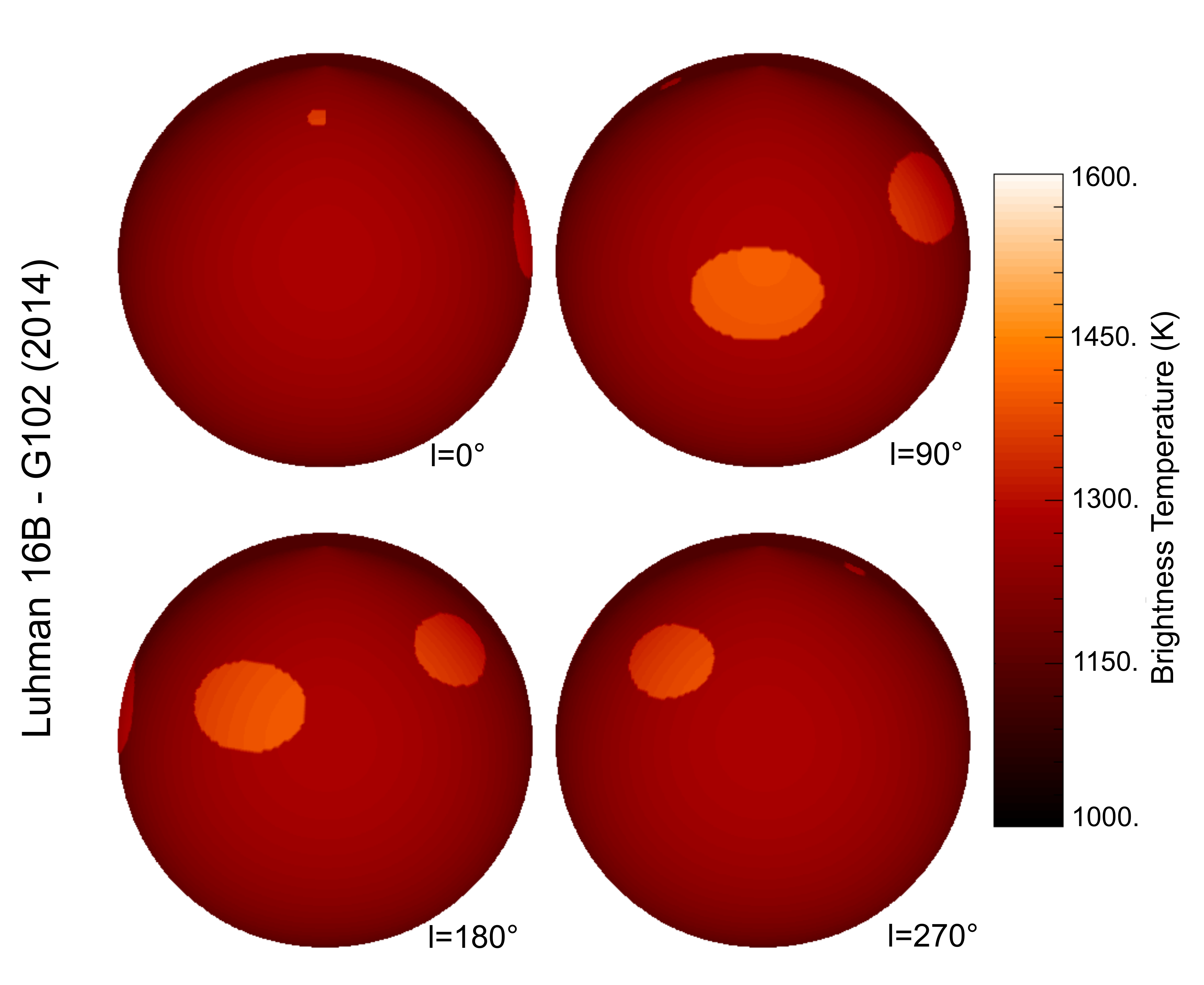}
\caption{Same as Fig.~\ref{fig:luhmanB_jmap}, but using the G102 grism light curve.}
\label{fig:luhmanB_jmap_epoch2}
\end{figure}

\subsection{Second epoch map}

Fig.~\ref{fig:luhmanB_jmap_epoch2} shows the brightness 
temperature map of Luhman 16B in the G102 grism. \textit{Aeolus} retrieved 4 spots 
(BIC 45 vs 61.7 for three spots) with (longitude, latitude) = 
($87^\circ\pm7^\circ$, $0^\circ\pm20^\circ$), ($154^\circ\pm18^\circ$,  
$28^\circ\pm12^\circ$), ($233^\circ\pm16^\circ$, $40^\circ\pm12^\circ$) and 
($355^\circ\pm15^\circ$, $63^\circ\pm8^\circ$)  and respective 
sizes of $37.63^\circ\pm0.92^\circ$, $39.56^\circ\pm1.89^\circ$, $35.87^\circ\pm2.10^\circ$ and 
$13.11^\circ\pm0.89^\circ$.
Assuming a background TOA temperature of 1280 K, the retrieved spots had a 
$\Delta T \sim 119\pm6$ K to the background TOA.  .

\begin{figure}
\centering
\includegraphics[height=60 mm]{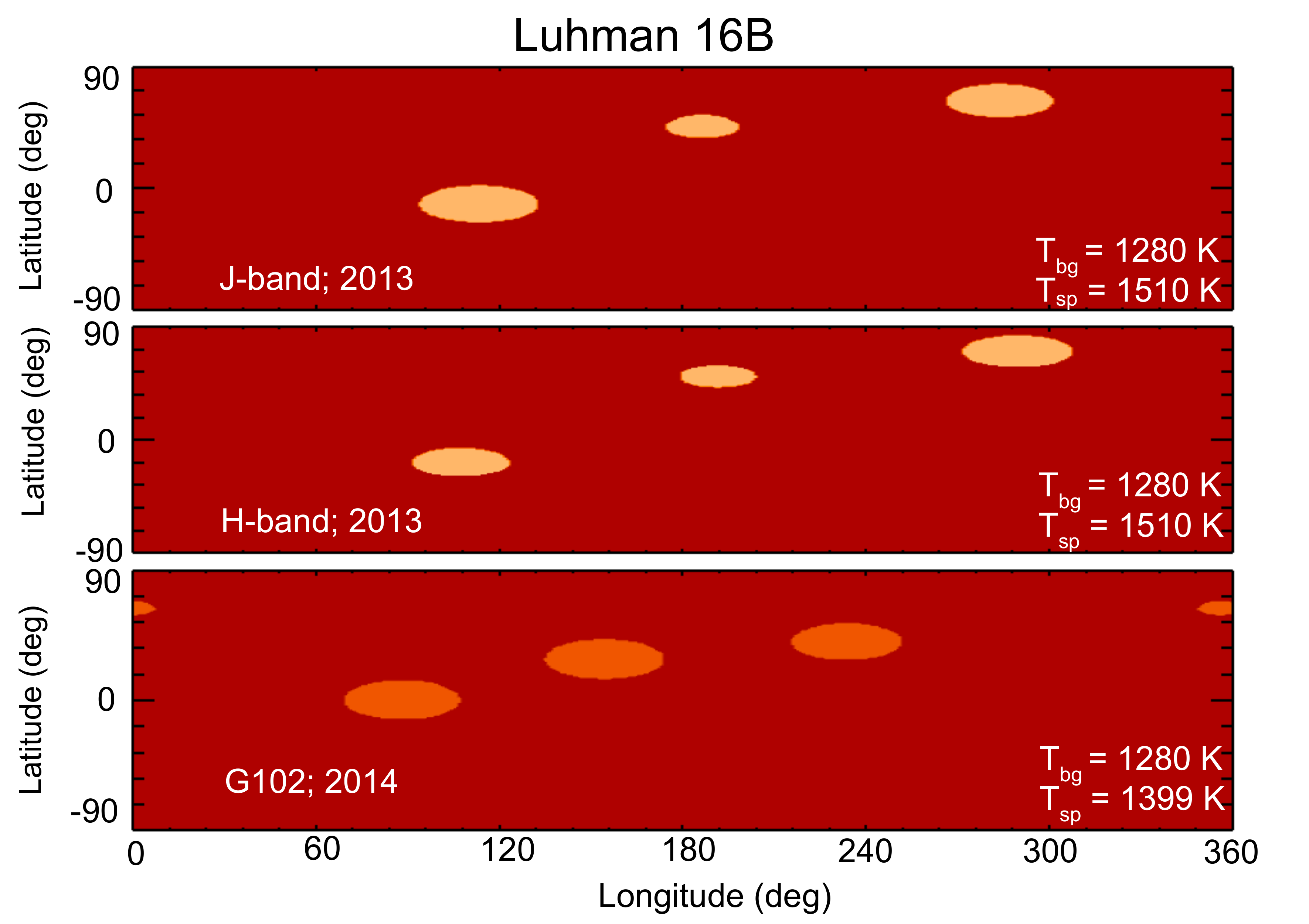} 
\caption{Luhman 16B maps, in an equirectangular projection, based on the J--band (top panel) and H--band (middle panel) light curves 
on 2013, and the G102 grism light curve on 2014 (bottom panel). The assumed background TOA brightness 
temperature ($T_\mathrm{bg}$) and the retrieved brightness temperature of the spots ($T_\mathrm{sp}$) is 
shown in every panel. }
\label{fig:luB_combi}
\end{figure}

In Table~\ref{table:obs2} we summarize 
the corresponding number of spots of the maps \textit{Aeolus} retrieved. 
For a direct comparison of the retrieved maps Fig.~\ref{fig:luB_combi} shows the 
2013 J-- (top panel) and H-- (middle panel) maps, and the 2014 G102 map (bottom panel) 
in an equirectangular projection. 

Note that the temperature differences retrieved for the spots in both epochs are 
comparable to the 200--300K temperature difference reported between the best--fit models of 
\citet[][]{buenzli15, buenzli15b}, and to the T$_\mathrm{hot}$-T$_\mathrm{cold}$ brightness temperature 
difference reported by \citet[][]{burgasser14}. These temperature differences are also comparable to the 
temperature difference \citet[][]{apai13} retrieved for the spots on two other brown dwarfs in the 
L/T transition SIMP0136 and 2M2139 (2MASSJ0136565+093347 and 
2MASSJ21392676+0220226 respectively).

\begin{table}[t]
\caption{Number of spots of the corresponding 
maps \textit{Aeolus} retrieved for the TOA of Luhman 16B per observational band and epoch.}
\centering
\resizebox{.3\textwidth}{!} {
\begin{tabular}{c c c }
\hline
\hline
Epoch & Band &   Number of spots\\
\hline
1 & J & 3 \\
 1&  H & 3 \\
 2 & G102 & 4\\ 
\hline
\end{tabular}
}
\label{table:obs2}
\end{table}

\section{Luhman 16A}\label{sect:luhm16a_maps}

\citet[][]{buenzli15b} observed variability in the atmosphere of  Luhman 16A, with its rotational 
light curve showing a peak--to--peak amplitude of $\sim$4.5\% (see bottom panel of Fig.~\ref{fig:luhmanB_lc}),  
allowing us to map Luhman 16A with \textit{Aeolus}.

We initially assumed a rotational period of 5 hr, following \citet[][]{buenzli15b}. As 
discussed in \citet[][]{crossfield14}, a similar rotational period for Luhman 16B and 16A would imply 
that the two brown dwarfs' rotational axes are misaligned. \textit{Aeolus} retrieved an 
inclination of $56^\circ\pm5^\circ$. 

\begin{figure}
\centering
\includegraphics[height=68mm]{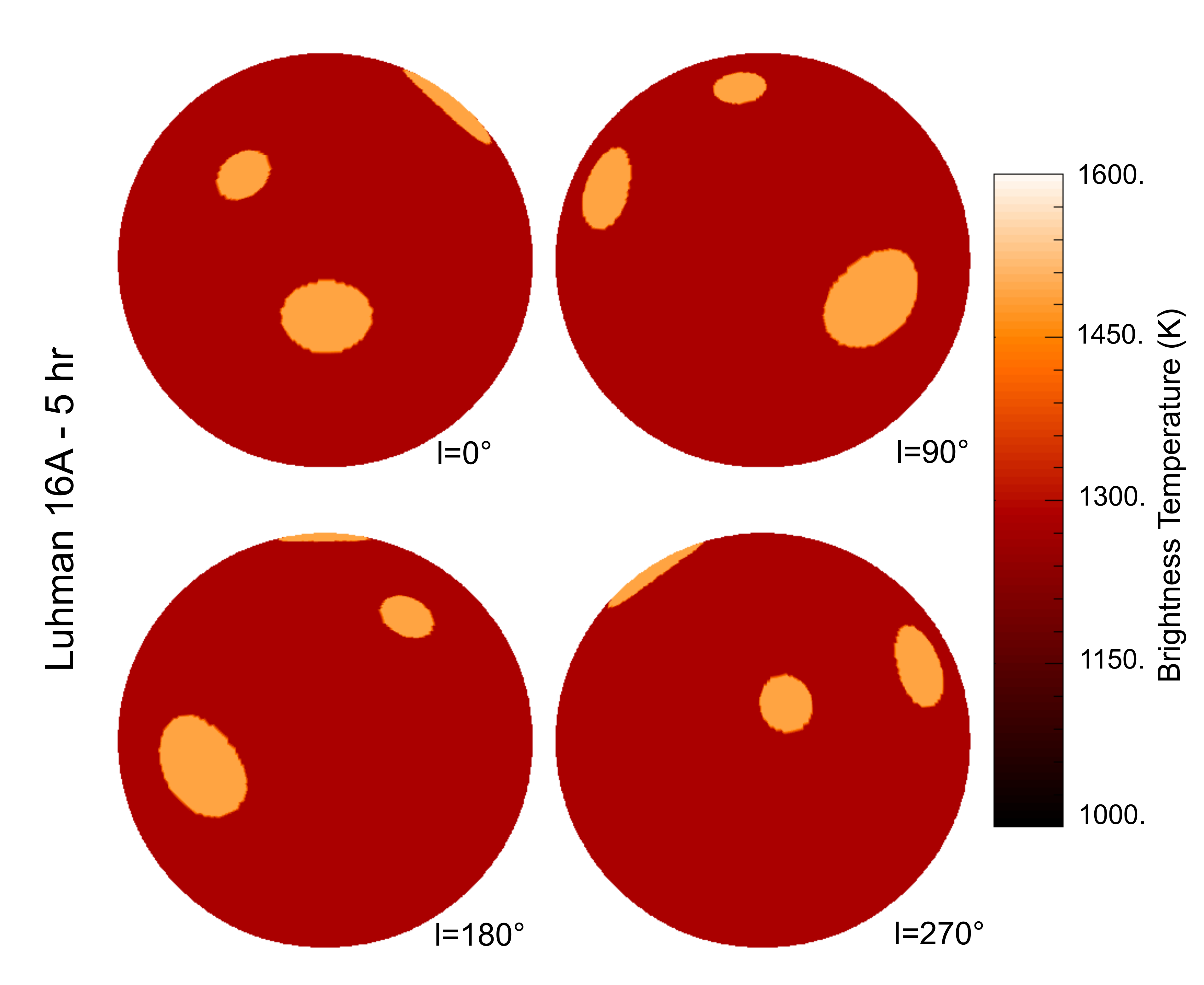}
\vspace{2pt}
\centering
\includegraphics[height=68mm]{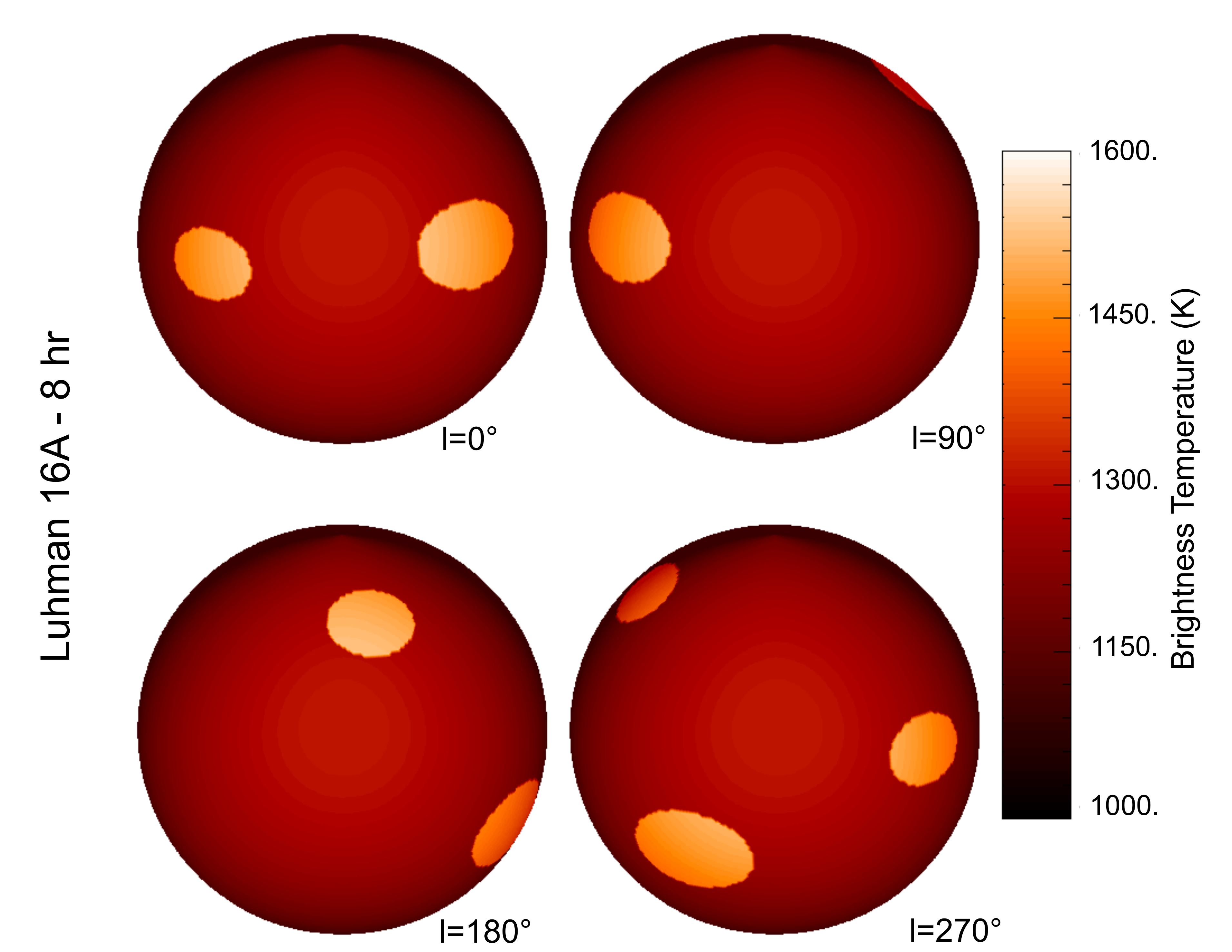}
\caption{Luhman 16A brightness temperature maps from applying 
\textit{Aeolus} on the HST/WFC3, G102 light curves of Fig.~\ref{fig:luhmanB_lc} (bottom panel), 
assuming a rotational period of 5 hr (top four panels), or 8 hr (top four panels). The maps 
are centered at $0^\circ$ of longitude (upper left map), $90^\circ$ of longitude (upper right map), 
$180^\circ$ of longitude (lower left map) and $270^\circ$ of longitude (lower right map). }
\label{fig:luhmanA_map}
\end{figure}

\begin{figure}
\centering
\includegraphics[height=64mm]{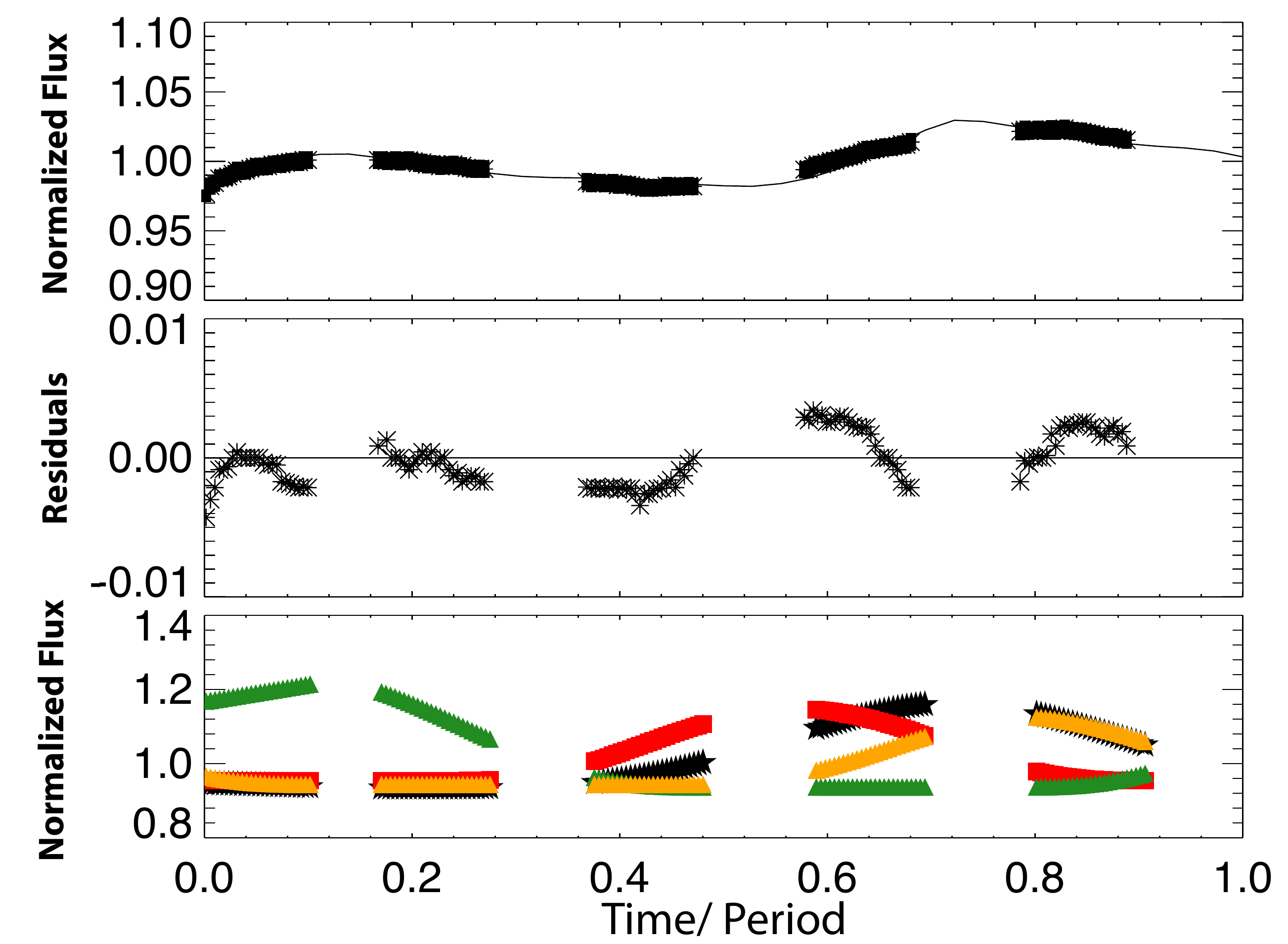}
\caption{Top panel: Normalized G102 light curve of Luhman 16A with a 8 hr rotational period (stars), 
with error bars and best fit \textit{Aeolus} curve (black solid line). Note that the error bars are smaller than the symbols. Middle panel: corresponding residuals. Bottom panel: Normalized light curves showing the contribution of each one of the four spots for our best-fit model.}
\label{fig:lon_histogram16A}
\end{figure}

Fig.~\ref{fig:luhmanA_map} (top four panels) shows the brightness temperature map of Luhman 16A 
assuming a rotational period of 5 hr. \textit{Aeolus} retrieved three spots 
(BIC$\sim$24.2 vs 37.5 for two spots) with 
(longitude, latitude) = ($115^\circ\pm8^\circ$, $32^\circ\pm10^\circ$), ($0^\circ\pm8^\circ$,  
$43^\circ\pm8^\circ$) and ($287^\circ\pm18^\circ$, $74^\circ\pm10^\circ$) and respective 
sizes of $37.72^\circ\pm6.96^\circ$, $35.65^\circ\pm1.10^\circ$ and $36.34^\circ\pm1.21^\circ$.
Assuming a background TOA temperature of 1310 K  \citep[][]{faherty14}, the retrieved spots had a 
temperature difference to the background TOA of $216\pm5$ K. 
In Fig.~\ref{fig:lon_histogram16A} we show the normalized G102 light curve of Luhman 16A (stars) 
with error bars and the best-fit \textit{Aeolus} curve (top panel), and the corresponding 
residuals (middle panel).

We then assumed a rotational period of 8 hr, following \citet[][]{mancini15}. \citet[][]{mancini15} 
performed spatially resolved, ground--based observations of Luhman 16AB over sixteen consecutive 
nights to estimate the rotational period of both components. Using a Gaussian process model 
and a MCMC analysis they retrieved a preferred period of 8 hr for Luhman 16A. However, 
they noted that a wide range of rotational periods are consistent with their data. 
\textit{Aeolus} retrieved an inclination of $18^\circ\pm8^\circ$, 
in agreement with the retrieved Luhman 16B inclination. As discussed in \citet[][]{crossfield14}, 
a different rotational period for Luhman 16A and B would allow the rotational axes of the two 
brown dwarfs to be aligned, as \textit{Aeolus} retrieved.  

Fig.~\ref{fig:luhmanA_map} (bottom four panels) shows the brightness temperature map of Luhman 16A 
assuming a rotational period of 8 hr.  \textit{Aeolus} retrieved four spots 
(BIC$\sim$21.6 vs 31.8 for three spots) with (longitude, latitude) = ($191.92^\circ\pm16.23^\circ$, 
$50^\circ\pm8^\circ$), ($243^\circ\pm10^\circ$,  $-24^\circ\pm10^\circ$), ($318^\circ\pm20^\circ$, 
$6^\circ\pm20^\circ$) and ($40^\circ\pm6^\circ$, $12^\circ\pm20^\circ$) and respective 
sizes of $38.67^\circ\pm0.57^\circ$, $39.44^\circ\pm1.63^\circ$, $28.88^\circ\pm1.16^\circ$ and $36.37^\circ\pm1.23^\circ$.
Assuming a background TOA temperature of 1310 K  \citep[][]{faherty14}, the retrieved spots had a 
temperature difference to the background TOA of $238\pm10$ K. 
Table~\ref{table:total_spots} summarizes the properties of all spots \textit{Aeolus} retrieved at the 
TOA of Luhman 16A and B, per observational band and epoch.

\begin{figure}
\centering
\includegraphics[height=58 mm]{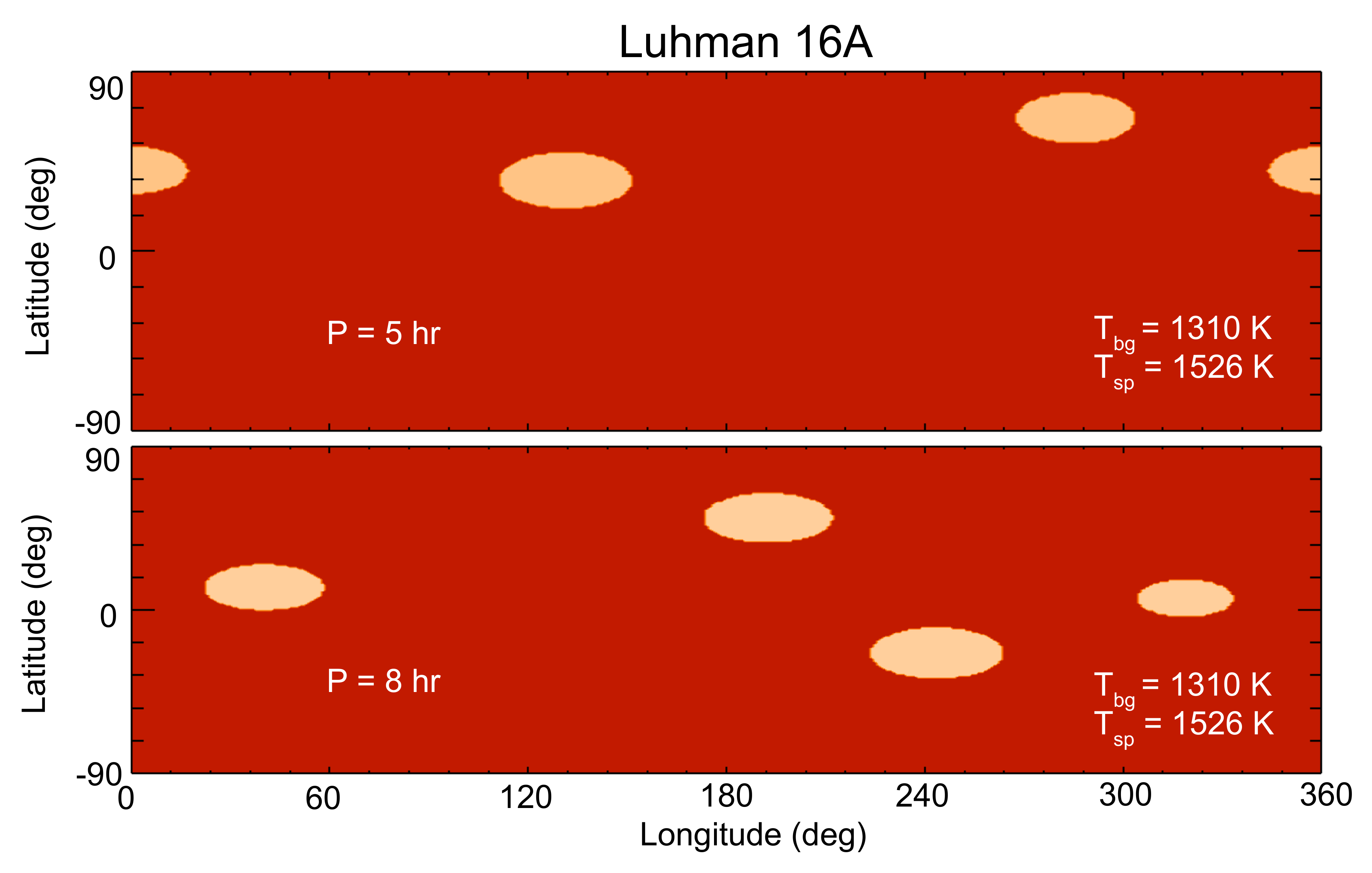} 
\caption{Luhman 16A maps assuming a rotational period of 5 hr (top panel), or 8 hr (bottom panel). }
\label{fig:luA_combi}
\end{figure}

\begin{table*}[t]
\caption{Properties of the spots \textit{Aeolus} retrieved at the TOA of 
Luhman 16A and B per observational band and epoch.}
\centering
\resizebox{.95\textwidth}{!} {
\begin{tabular}{c c c c c c c c}
\hline
\hline
Target & Period (hr) & Epoch & Band & Longitdue & Latitude & Size & Temperature Contrast \\
\hline
Luhman 16A & 5 & 2 & G102 & $115^\circ\pm8^\circ$ & $32^\circ\pm10^\circ$ & $37.72^\circ\pm6.96^\circ$ & $216\pm5$ K\\
          &   &  &  &  $0^\circ\pm8^\circ$ &   $43^\circ\pm8^\circ$  & $35.65^\circ\pm1.10^\circ$ &$216\pm5$ K  \\
          &  &  &  & $287^\circ\pm18^\circ$ &  $74^\circ\pm10^\circ$ & $36.34^\circ\pm1.21^\circ$ &$216\pm5$ K \\
Luhman 16A & 8 & 2 & G102 & $191.92^\circ\pm16.23^\circ$ & $50^\circ\pm8^\circ$ & $38.67^\circ\pm0.57^\circ$ & $238\pm10$ K\\
          &   &  &  &  $243^\circ\pm10^\circ$ &  $-24^\circ\pm10^\circ$  &  $39.44^\circ\pm1.63^\circ$ &  $238\pm10$ K\\ 
          &  &  &  & $40^\circ\pm6^\circ$ & $12^\circ\pm20^\circ$ & $36.37^\circ\pm1.23^\circ$ & $238\pm10$ K\\
          &  &  &  & $318^\circ\pm20^\circ$ & $6^\circ\pm20^\circ$ & $28.88^\circ\pm1.16^\circ$ & $238\pm10$ K\\
Luhman 16B & 5.05 & 1 & J & $113.47^\circ\pm6.46^\circ$ & $-20^\circ\pm12^\circ$ & $38.89^\circ\pm0.80^\circ$ & $231\pm16$ K \\
          &  &  &  &$186.08^\circ\pm3.65^\circ$ & $45^\circ\pm12^\circ$ & $24.45^\circ\pm1.46^\circ$ & $231\pm16$ K\\
          &  &  &  &$283^\circ\pm13^\circ$ & $72^\circ\pm10^\circ$ &  $35.54^\circ\pm1.67^\circ$ & $231\pm16$ K\\
Luhman 16B & 5.05 & 1 & H & $107.27^\circ\pm5.48^\circ$ & $-18^\circ\pm8^\circ$ & $32.42^\circ\pm0.80^\circ$ & $211\pm16$ K \\
          &  &  &  &$190.91^\circ\pm4.83^\circ$ & $51^\circ\pm20^\circ$ & $25.02^\circ\pm1.13^\circ$ & $211\pm16$ K\\
          &  &  &  &$288.6^\circ\pm9.4^\circ$ & $74^\circ\pm6^\circ$ &  $36.30^\circ\pm1.18^\circ$ & $211\pm16$ K\\
Luhman 16B & 5.05 & 2 & G102 & $87^\circ\pm7^\circ$ & $0^\circ\pm20^\circ$ & $37.63^\circ\pm0.92^\circ$ & $119\pm6$ K\\
          &  &  &  & $154^\circ\pm18^\circ$ &  $28^\circ\pm12^\circ$ & $39.56^\circ\pm1.89^\circ$ & $119\pm6$ K\\
          &  &  &  & $233^\circ\pm16^\circ$ & $40^\circ\pm12^\circ$ & $35.87^\circ\pm2.10^\circ$ & $119\pm6$ K\\
          &  &  &  & $355^\circ\pm15^\circ$ & $63^\circ\pm8^\circ$ & $13.11^\circ\pm0.89^\circ$ & $119\pm6$ K\\
 \hline
\end{tabular}
}
\label{table:total_spots}
\end{table*}

\section{Discussion}~\label{sect:discussion}

\subsection{Comparison of \textit{Aeolus} map with previously published Luhman 16B map.}

\begin{figure}
\centering
\includegraphics[height=68mm]{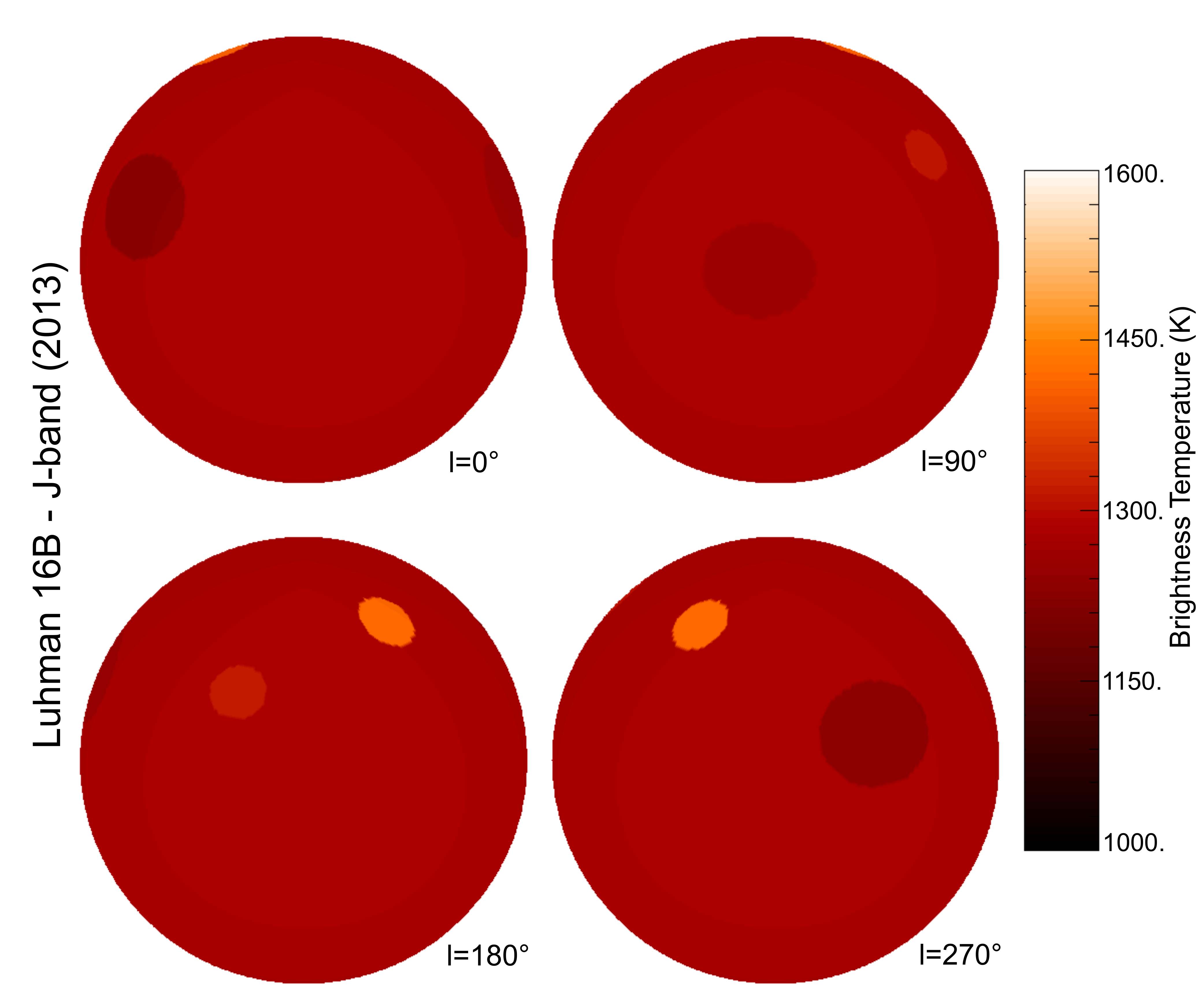}
\vspace{2pt}
\centering
\includegraphics[height=50mm]{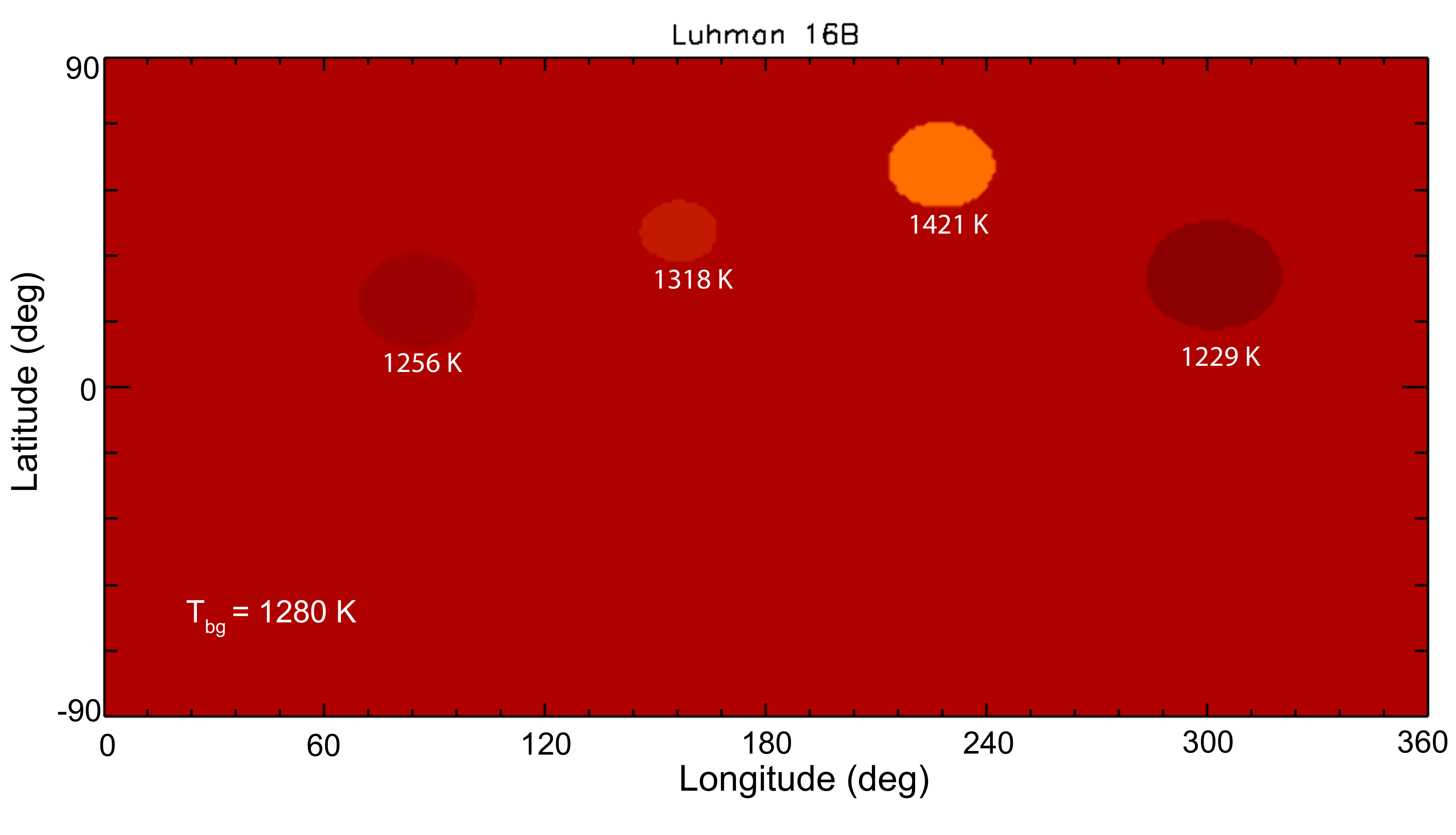}
\vspace{2pt}
\centering
\includegraphics[height=60mm]{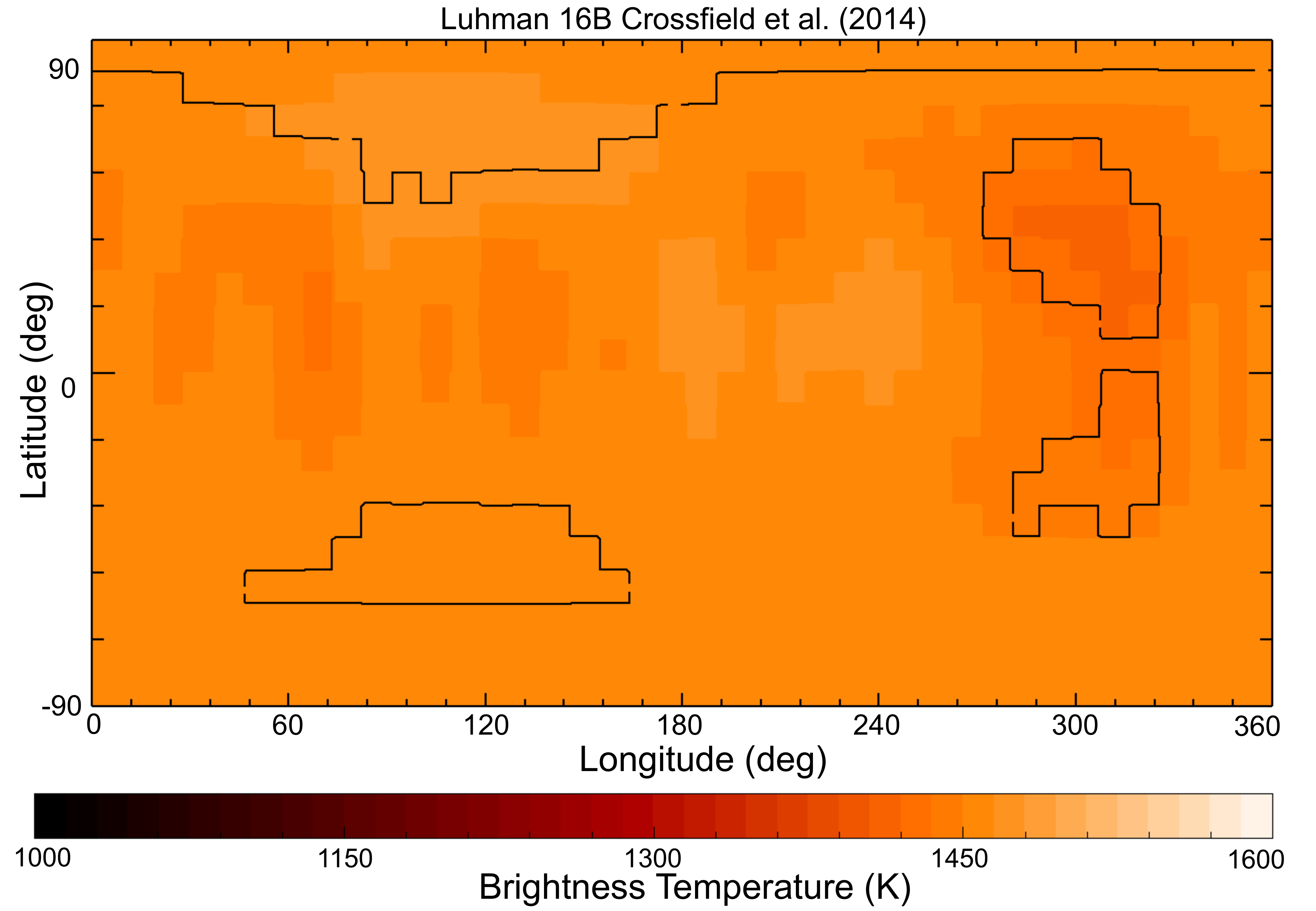}
\caption{Top panel: Same as in Fig.~\ref{fig:luhmanB_jmap} (top panels), but allowing \textit{Aeolus} 
to fit the brightness temperature (i.e., contrast ratio) of every spot independently of the others. 
Middle panel: Same map as top panel, but in an equirectangular projection. 
Bottom panel: \citet[][]{crossfield14} map in an equirectangular projection. Black contour lines 
show the features with the highest signal--to--noise ratio ($\geq 3$): a polar, brighter than the background TOA 
spot and a mid-latitude, darker than the background TOA spot.}
\label{fig:luhmanB_mulit_map}
\end{figure}

\citet[][]{crossfield14} mapped Luhman 16B  using Doppler imaging and found a 
complex map, with both brighter and darker than the background TOA cloud patches (see 
bottom panel of Fig.~\ref{fig:luhmanB_mulit_map}). This contradicts 
the PCA analysis of \citet[][]{buenzli15, buenzli15b} that found that only one kind of cloud patches/spots 
is necessary to explain the variability of the observed spectra. Given the night--to--night 
variability of Luhman 16B's light curves one could argue that the observed difference could be due to the 
$\sim$0.5 year between the \citet[][]{crossfield14} and \citet[][]{buenzli15} observations. 
However, the \citet[][]{buenzli15} and \citet[][]{buenzli15b} observations were taken $\sim$1 year apart and the PCA 
analysis of the two datasets were in agreement. \citet[][]{crossfield14} observed Luhman 16B in 
the CO absorption lines in the K--band, probing lower pressure levels in the Luhman 16B atmosphere than the G141 and 
G102 grism observations of \citet[][]{buenzli15, buenzli15b} ($\sim$0.94 bar versus $\sim$2.7 bar, 
see Fig.~\ref{fig:contr_funct}). The different pressure levels probed could explain the difference between the 
mixed--spot map of \citet[][]{crossfield14} and the single--temperature--spot maps implied by the 
PCA analysis of \citet[][]{buenzli15, buenzli15b}. However, this could imply that different mechanisms 
rule the formation of spots in the deeper and upper atmosphere (blocking or allowing the formation of 
brighter and darker spots, respectively). Note that in a Doppler imaging map brighter/darker spots could 
also be caused by variations in the atmospheric abundances across the TOA. Assuming that some 
of these features could indicate abundance variations, rather than cloud, heterogeneities could explain 
the difference between the PCA analysis and the \citet[][]{crossfield14} mixed-spot map. However, 
HST spectral mapping by \citet[][]{buenzli15, buenzli15b} showed no evidence for abundance variations 
in the G102 and G141 grism. 

For a direct comparison of our maps with the \citet[][]{crossfield14} map (hereafter C14 map)
we again applied \text{Aeolus} on the J--band light curve of Fig.~\ref{fig:luhmanB_lc}, 
this time allowing \textit{Aeolus} to fit the contrast ratio of every spot independently of the others. 
\textit{Aeolus} retrieved four spots (BIC$\sim$38.9) with (longitude, latitude) = ($84.72^\circ\pm8.83^\circ$, 
$23^\circ\pm8^\circ$), ($155.87^\circ\pm8.47^\circ$,  $42^\circ\pm9^\circ$), ($227.3^\circ\pm8.7^\circ$, 
$59^\circ\pm6^\circ$) and ($301.28^\circ\pm8.08^\circ$, $30^\circ\pm11^\circ$) and respective 
sizes of $31.99^\circ\pm1.19^\circ$, $20.87^\circ\pm1.12^\circ$, $28.85^\circ\pm1.16^\circ$ 
and $36.92^\circ\pm1.14^\circ$. Assuming a background TOA temperature of 1280 K, 
the retrieved spots had a $\Delta T \sim$  $-23\pm6$ K, $38\pm6$ K, $141\pm6$ K and 
$-51\pm6$ K to the background TOA. Note that the multiple-component model has a larger BIC than our 
single--component model ($\sim$39 vs $\sim$26), so the latter would be preferred (by \textit{Aeolus}) over the former.

Fig.~\ref{fig:luhmanB_mulit_map}  (top and middle panels) shows the brightness temperature maps 
of Luhman 16B allowing \textit{Aeolus} to fit the contrast ratio of all (possible) spots independently 
of the others. The Doppler imaging map of  \citet[][]{crossfield14} is also shown for comparison 
(bottom panel). As in \citet[][]{crossfield14}, \textit{Aeolus} found a mixture of darker and brighter 
than the background TOA spots. We compared the spots \textit{Aeolus} retrieved with the highest 
signal--to--noise--ratio features of the C14 map (see Fig.~\ref{fig:luhmanB_mulit_map}). Note though, 
that the $\sim$0.5 year between the \citet[][]{crossfield14} and \citet[][]{buenzli15} observations, in combination 
with the observed night--to--night evolution of Luhman 16B's light curves (and thus TOA structure), indicates 
that, probably, our maps probed very different features from the C14 map.

An interesting result is that \textit{Aeolus} retrieved a very large, dark 
spot (the largest, and darkest spot in the \textit{Aeolus} map) in agreement with the C14 map, 
even though the latter map probed higher altitudes in the atmosphere 
($\sim$0.94 bar versus $\sim$2.7 bar of our map). Both the 
\textit{Aeolus}' and the C14 map's darkest spots lay at low latitudes. The 
spot \textit{Aeolus} retrieved was $\sim51$ K cooler than the background TOA and spanned 
$\sim37^\circ$ of longitude, while the spot of the C14 map was $\sim40$ K cooler than the background TOA (assuming 
a background temperature of $\sim$1450 as in \citet[][]{crossfield14}) and 
spanned $\sim42^\circ-48^\circ$ of longitude. 
The brightest spot \textit{Aeolus} retrieved was $\sim141$ K hotter than the background TOA 
and lay at mid latitudes, while the brightest spot of the C14 map was $\sim26$ K hotter than the background 
TOA and lay at high latitudes. Finally, the brightest and darkest spots \textit{Aeolus} 
retrieved lay $\sim74^\circ$ of longitude apart, while the brightest and 
darkest spots of the C14 map lay $\sim190^\circ$ of longitude apart.

A notable difference between the \textit{Aeolus} and the C14 maps is that 
the latter contains latitudinally extensive cloud patches, while the former contains narrower spots. 
The latitudinal extensive patches are intrinsic to the Doppler imaging 
maps, since they appear even when the input maps contain latitudinally narrow spots 
(compare, e.g., the input map and the retrieved maps of \citet[][]{unruh95}, or 
the input and retrieved maps of \citet[][]{crossfield14} [their Fig. 5]). 
The latitudinal extensive cloud patches of the C14 maps thus, 
do not contradict \textit{Aeolus}' latitudinally narrower spots. 

Another difference between the C14 and the \textit{Aeolus} maps is that 
the former shows more features than the latter. 
In \citet[][]{karalidi15} we showed that \textit{Aeolus} could not detect small scale 
features of the TOA ($\lesssim 10^\circ$). However, most of the features in the C14 map are extensive 
enough for \textit{Aeolus} to have mapped them. Most of the features in the C14 map 
have a very low signal--to--noise--ratio making their detection ambiguous. 
Some of these features could be artifacts of the Doppler imaging due to, e.g., a slightly offset 
assumed limb darkening or inclination \citep[see, e.g.,][]{vogt87, unruh95}. Additionally, given that 
Doppler imaging is also sensitive to abundance variations across the atmosphere, these features 
could indicate abundance, rather than cloud, heterogeneities. Finally, note again that 
the time elapsed between the two observations (\citet[][]{crossfield14} and \citet[][]{buenzli15}) indicates 
that, probably, our maps probed very different features from the C14 map.

\subsection{Implications of \textit{Aeolus} maps for the atmospheres of Luhman 16A \& B.}

\subsubsection{Wind speeds and characteristic timescales.}

Assuming that the maximum spot size retrieved by \textit{Aeolus} is defined by the atmospheric 
jet size, we can follow a Rhines--length--based argument as in \citet[][]{apai13, burgasser14} 
and \citet[][]{karalidi15} to constrain the wind speed on Luhman 16A and B. 
We assume that our maps are accurate, and that the retrieved spots in the TOA are uniform.  
We additionally assume that Luhman 16A's and B's radii are equal to one 
Jupiter radius (evolutionary models suggest that the radius of Luhman 16B, assuming an age 
between 0.5 Gyr and 5 Gyr, is $\sim1.0\pm0.2$ R$_\mathrm{Jup}$, see \citet[][]{saumon08, burrows11})
and we use the maximum spot size \textit{Aeolus} retrieved to calculate the wind speed 
in the atmospheres of Luhman 16A and B as: $u_\mathrm{wind}\sim \Omega R/s^2$, 
where $\Omega$ is the angular velocity, $R=1\times R_\mathrm{Jup}$ and $s$ is the maximum 
spot size. 
For Luhman 16A this implies a wind speed $u_\mathrm{wind}\sim$ 602$\pm$49 
or 891$\pm$58 m/s, depending on the assumed rotation rate (8 hr or 5 hr, and assuming 
a $\delta p\sim$ 0.1 hr similar to Luhman 16B), while for Luhman 16B this implies a 
$u_\mathrm{wind}\sim$ 813$\pm$55 m/s (934$\pm$37 m/s) based on the H-- (J--) band map, 
and $\sim$ 968$\pm$94 m/s based on the G102 map. These speeds are higher than the 
wind speeds of the giant planets of our Solar system, but are lower, or comparable to 
the wind speeds reported for highly irradiated hot Jupiters by, e.g., \citet[][]{louden15, colon12} 
and \citet[][]{snellen10}. Previous calculations of the wind speed on Luhman 16B by 
\citet[][]{burgasser14} predicted wind speeds between 1600 m/sec and 3400 m/sec, 
assuming a background TOA temperature of 1510K and a spot--temperature of 
1700K$<$T$<$1900K. These speeds are higher than the ones \textit{Aeolus} retrieved. 
However, since these speeds are upper limits our results do not contradict, but rather 
further constrain the results of \citet[][]{burgasser14}.

We performed a back--of--the--envelope calculation of the speed of sound, $c_s$, 
on Luhman 16A and B. H$_2$ is the major constituent of both  
brown dwarfs' and Jupiter's atmospheres. We took thus into account the variation of the specific 
heat ratio of H$_2$ between Jupiter's $\sim$165 K and Luhman 16A and B $\sim$1300 K, as well as the 
variation between these atmospheres' relative molecular mass. Assuming that 
$c_{s-\textrm{Jupiter}}\sim$1,000 m/sec at 1.5 bar \citep[][]{lorenz98}, we found that in Luhman 
16A and B $c_s$ $\sim$2800 m/sec. This implies that the wind speeds we retrieved using 
\textit{Aeolus} and a Rhines--length--based argument are subsonic, and thus plausible for 
Luhman 16A and B atmospheres.

Using the retrieved wind and sound speeds for the atmospheres of Luhman 16A and B 
we calculated the minimum timescales associated with wind-- and density--wave-- driven 
changes in these atmospheres. Assuming that Luhman 16A's and B's 
radii are equal to Jupiter's radius and setting $t_s\sim R/c_s$ we find 
$t_s\sim$6.94 hr, i.e., $\sim$1.37 Luhman 16B rotations and $\sim$1.37 or $\sim$0.87 Luhman 16A  
rotations, depending on the assumed rotational period (5 hr or 8 hr respectively). 
The advection timescale $t_w\sim R/u_\mathrm{wind}$ of Luhman 16A's atmosphere 
is $t_w\sim$ 32.26 hr (23.65 hr) or 4 (4.68) Luhman 16A 
rotations, depending on the assumed rotational period. For Luhman 16B $t_w\sim$ 23.89 hr (20.79 hr, or 
20.06 hr) or 4.7 (4.12, or 3.97) rotations based on the H--band (J--band, or G102 grism) map. 
If we define a timescale $t_h$ as the time required for a spot to be horizontally displaced by 
a full spot--sized length ($\sim 40^\circ$ or $\sim 0.7 R$) due to the atmospheric wind, 
we get:  $t_h\sim$ 22.6 hr (16.6 hr) or 2.8 (3.3) Luhman 16A rotations, and 
 $t_h\sim$ 16.7 hr (14.6 hr, or 14 hr) or 3.3 (2.9, or 2.8) Luhman 16B rotations.
Space and ground--based observations of Luhman 16B show evolution of its light curve 
within a couple of rotations, indicating that the TOA structure of Luhman 16B changes 
in timescales of 1--5 rotations (see, e.g., \citet[][]{buenzli15, buenzli15b, mancini15}). 
This is comparable to the timescales we inferred here for  density--wave-- and 
horizontal wind-- driven changes in the atmosphere of Luhman 16B.


\subsubsection{Luhman 16B's evolving cloud structures.}

\citet[][]{apai13} and \citet[][]{buenzli14b, buenzli15} showed that the observed variability of 
light curves of brown dwarfs in the L/T transition is not 
caused by cloud clearings, but by variations in the optical thickness of clouds across the 
TOA. A similar behavior is observed for the giant, cloud covered, planets of our Solar System. For example, 
\citet[][]{karalidi15} showed that the disk--integrated light curves of Jupiter in the F275W and F763M 
are comparable to observed light curves of brown dwarfs, and are dominated by two 
distinct cloud features: the Great Red Spot (a towering cloud structure in the Jovian atmosphere) 
and a major 5 $\mu$m Hot Spot (a region of thinner clouds that allows us to see deeper cloud layers in the Jovian 
atmosphere). Recently, \citet[][]{simon15} acquired K2 observations of Neptune over 49 days and showed that 
the disk--integrated light curve of Neptune is dominated by a few distinct cloud features. 
Using the K2 observations in combination with Hubble Space Telescope and Keck imaging 
\citet[][]{simon15} proposed that a combination of stable, large--scale cloud features and 
smaller, short--lived cloud features defines the TOA structure and the observed light curve 
variability. A similar combination could be responsible for the observed variability in 
brown dwarf light curves. 
The best--fit maps of \textit{Aeolus} for Luhman 16A and B have spots that are 
hotter than the background TOA. This could be, for example, the case when due to thinner 
clouds we see deeper, hotter layers of the atmosphere. These areas of thinner clouds could 
be caused by convective downdrafts, 
in a similar way that Jupiter's 5 $\mu$m Hot Spots are thought to be formed 
\citep[see, ][]{showman98,showman00}.

\begin{figure}
\centering
\includegraphics[height=60mm]{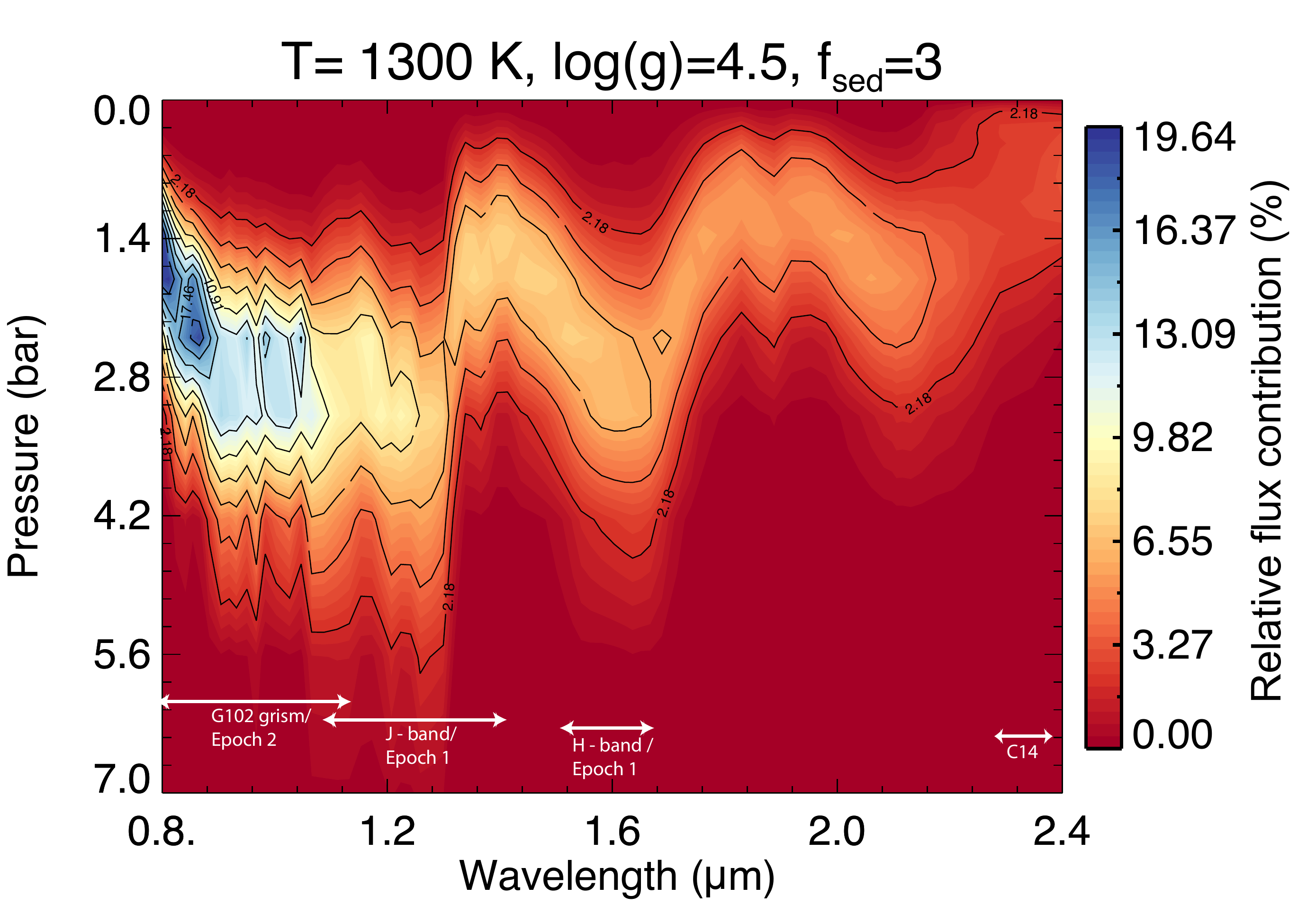}
\caption{Relative flux contribution of every pressure level in the best-fit model 
atmosphere of \citet[][]{buenzli15} as a function of wavelength. The wavelength 
ranges observed in the two observational epochs used in this paper, as well as in \citet[][]{crossfield14} 
(C14), are shown with white arrows.}
\label{fig:contr_funct}
\end{figure}

Using the contribution function of a T2 dwarf for Luhman 16B we note that the J--band 
and the G102 grism probe approximately the same pressure ranges (see Fig.~\ref{fig:contr_funct}). When 
comparing the maps of the two epochs one can thus deduce the evolution of the pressure layer over 
the $\sim$1800 rotations between the two observations. The maps show 
considerable difference between the two epochs (see Fig.~\ref{fig:luB_combi}). 
However, we caution that the spots of the second 
epoch are probably not the same as the ones of the first epoch. Luhman 16B light curves exhibit a 
significant evolution from one rotational period to the next. Assume that the observed 
evolution over the $\sim$9,000 hr ($\sim$1800 rotations) between the two observational epochs 
is solely due to the longitudinal shift of spots due to winds and possible appearance of new spots.  
If we then assume a constant wind speed of $u_\mathrm{wind}\sim$934 m/s 
being responsible for the displacement of the spots of epoch 1, in epoch 2 the spots 
would be displaced by $\sim68^\circ$ of longitude. This is a $\sim$38\% to $\sim$75\% larger 
displacement than what we observe in our maps, implying that more mechanisms 
are responsible for our light curve evolution. 


Space and ground--based observations of Luhman 16B show a rapid evolution of 
the observational light curves, indicating a rapidly changing atmosphere. During 
the $\sim$1800 rotations that separate the two epochs of the observations used in this paper, 
Luhman 16B's TOA structure should have varied multiple times. Indeed, 
\citet[][]{mancini15} ground--based observations of Luhman 16B between 2014 April 19 and 2014
July 16 showed that Luhman 16B's light curve evolved from one night to the next (see their 
Fig. 6). Note that the \citet[][]{mancini15} observations were made in the $i+z$ bands, 
which, shortward of $0.8 \mu$m, probe higher altitudes of Luhamn 16B's atmosphere 
than the G102 band. 

It is interesting to estimate the TOA map of Luhman 16B between the two observational 
epochs of this paper using the \citet[][]{mancini15} light curves (their Fig. 6, hereafter MLC). 
We were not interested in an accurate mapping, but rather a sketch of the TOA 
structure. Thus, we did not apply \textit{Aeolus} on the MLC but instead we performed 
a visual comparison of the MLC with light curves \textit{Aeolus} produced, assuming that 
all (possible) spots are hotter than the background TOA by $\sim200$~K (see Fig.~\ref{fig:luB_combi}). 
The best--fit maps included three--or--more spots, whose location on the TOA varied 
from one light curve (night) to the next. For example a visual comparison of \textit{Aeolus} 
light curves with the ``MDJ-56770'' light curve gives a best-fit for three spots, for the 
``MDJ-56775'' light curve for four spots, and for the ``MDJ-56778'' light curve for three spots. 
Given that the MLC were separated 
by as little as $\sim$5 rotations from each other, the variability 
of our best--fit maps indicates a highly variable atmosphere. 
In the future, the acquisition 
of continuous, multi--rotation observations of Luhman 16B will be of great 
interest. Applying \textit{Aeolus} on the light curves of such observations, 
we will be able to continuously map the variability of Luhman 16B's 
TOA and provide feedback to General Circulation and Radiative Transfer 
models to help understand the mechanisms that rule the observed variability. 

\subsubsection{The Possible Persistent Cloud Structure PPCS-1}

An inspection of the J-- (or H--) band light curve of epoch 1 
and the G102 grism light curve of epoch 2 showed the existence of a similar trough around a rotational 
phase of 0.6. The shape of these troughs was similar even though the two observational 
epochs were separated by $\sim1800$ rotations (see Fig.~\ref{fig:luhmanB_lc}). 
The similarity of the troughs in these light curves hints to the existence of a 
similar feature in the TOA of Luhman 16B in the two epochs. 
We hereafter refer to this feature as: Possible Persistent Cloud Structure PPCS-1. 
We observed a similar trough in some of the light curves of \citet[][]{gillon13} (observed in the 
$I+z$ TRAPPIST filter), while none of the \citet[][]{mancini15} light curves (observed in the $i+z$ 
Danish 1.54m long--pass filter) contained 
a trough that matched the PPCS-1.

\begin{figure}
\centering
\includegraphics[height=60mm]{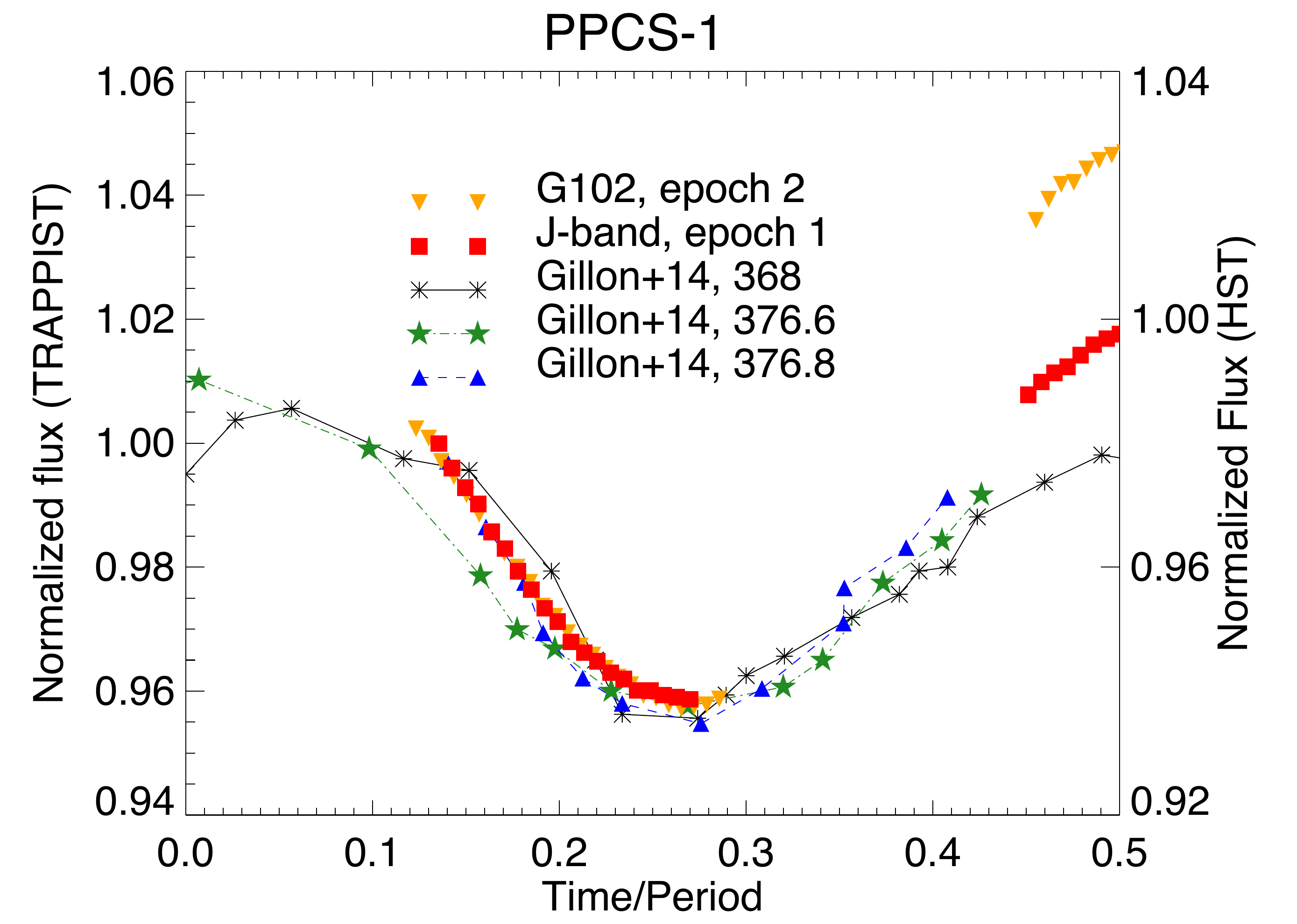}
\caption{Luhman 16B light curves observed hundreds of rotations from each other exhibit a similar 
trough, indicating the existence of a similar feature in the TOA of Luhman 16B (PPCS-1). Note that 
we shifted the light curves so that the trough of PPCS-1 match.}
\label{fig:feature_A}
\end{figure}

Fig.~\ref{fig:feature_A} shows the PPCS-1 in the two epochs of HST light curves used 
in this paper, and the matching PPCS-1 of \citet[][]{gillon13} light curves. Note that the phase 
(time/period) of the light curves is altered in comparison to Fig.~\ref{fig:luhmanB_lc}, and that 
we shifted the light curves so that the troughs of PPCS-1 match. The 
\citet[][]{gillon13} observations took place $\sim2800$ rotations before the epoch 1, and 
$\sim4600$ rotations before the epoch 2 observations used in this paper. The existence of 
PPCS-1 in these five light curves separated by tens of hundreds of rotations is intriguing. 
Note that the \citet[][]{gillon13} observations used the $I+z$ filter of the TRAPPIST telescope, 
and probe higher (similar) altitude levels in the atmosphere of Luhman 16B than our G102 grism light curve 
shortwards (longwards) of $\sim0.8 \mu$m. 

We performed a back--of--the--envelope calculation of the orbital period of an exoplanet responsible for 
PPCS-1. Taking into account the observed J--band ``transit'' depth and duration, and assuming 
a circular orbit with an impact parameter $b=0$ (the planet transits 
through the center of the Luhman 16B's disk),  we calculated an 
orbital period of $\sim101$ hr [a more accurate fit of 
the PPCS-1 to possible exoplanet orbits gave a best-fit for periods of $\gtrsim500$ days 
(Ben W. P. Lew, private communication)].
We thus ruled out the possibility that PPCS-1 was due to an exoplanet, since the 
PPCS-1 was visible in two successive Luhman 16B rotations in the \citet[][]{gillon13} light curves 
(see the ``Gillon+14, 376.6'' and ``Gillon+14, 376.68'' PPCS-1 in Fig.~\ref{fig:feature_A}).  

Assuming the PPCS-1 is due to a TOA feature this could indicate either the existence of 
a stable formation (like the Great Red Spot of Jupiter) that periodically reappears at the TOA, 
or similar formations that appear and disappear at different times. In the case of a stable formation 
variations in other, unrelated cloud structures could occasionally wash out the modulations, thus 
explaining why the PPCS-1 is only detected in some, but not other, light curves. In the case of similar 
formations appearing at different observation times this could imply a possible preferred size for 
cloud structures in the TOA of Luhman 16B. Longer, multi-wavelength observations of Luhman 16B 
over (partially) continuous rotations could help clarify the nature of this feature.

\section{Conclusions}~\label{sect:concl}
We presented the first map of Luhman 16A and maps of 
 two epochs of Luhman 16B. \textit{Aeolus} constrained the inclination of 
 Luhman 16A to $18^\circ\pm8^\circ$ or $56^\circ\pm5^\circ$, depending on the 
 assumed rotational period, and $26^\circ\pm8^\circ$ for Luhman 16B. 
 
 In agreement with the complexity of the observational light curves \textit{Aeolus} 
 retrieved complex top--of--the--atmosphere cloud structures for both Luhman 16A and 
 16B, with a surface spot coverage of 19\% to 32\% (depending on the assumed rotational period), 
 and $\sim$21\% (2013) to $\sim$39\% (2014) respectively. 
 
We compared our Luhman 16B maps with the only previously published map of 
\citet[][]{crossfield14}. Using the principal component analysis results of \citet[][]{buenzli15, buenzli15b}  
\textit{Aeolus} retrieved 
hotter than the background TOA spots for both observational epochs, unlike the 
hotter and cooler than the background TOA spots mixture that \citet[][]{crossfield14} map 
showed. Relaxing the PCA induced constraints \textit{Aeolus} fit a mixture of 
hotter and cooler (than the background TOA) spots, in agreement with the 
\citet[][]{crossfield14} map. Interestingly, the largest spot \textit{Aeolus} retrieved 
was the coolest spot and lay at low latitudes, in agreement with the \citet[][]{crossfield14} map, 
even though the latter probed higher altitudes in the atmosphere of Luhman 16B.
However, the BIC of the mixed--spot solution was larger than that 
of hotter--only spot solution, making the latter the preferred \textit{Aeolus} map.

Finally, we reported the detection of a feature (PPCS-1) that reappeared in light curves of Luhman 16B that are 
separated by tens of hundreds of rotations from each other. We excluded the possibility 
that this feature is due to an exoplanet and speculated that it is related to TOA structures of 
Luhman 16B. \\

 \acknowledgements
This work is part of the Spitzer Cycle-9 Exploration Program
Extrasolar Storms (program No. 90063). Support for this work was provided by 
NASA through an award issued by JPL/Caltech.
Support for Program number 12314 was provided by NASA
through a grant from the Space Telescope Science Institute,
which is operated by the Association of Universities for
Research in Astronomy, Incorporated, under NASA contract
NAS5-26555. An allocation of computer time from the UA Research Computing High Performance 
Computing (HTC) and High Throughput Computing (HTC) at the University of Arizona 
is gratefully acknowledged. This study, in part, is based on observations 
made with the NASA/ ESA Hubble Space Telescope, obtained at the Space 
Telescope Science Institute, which is operated by the Association of Universities 
for Research in Astronomy, Inc., under NASA contract NAS 5Ð26555. D. Apai 
acknowledges support by the National Aeronautics and Space Administration under Agreement
No. NNX15AD94G for the program Earths in Other Solar Systems. 
We thank I.J.M. Crossfield for providing us with the previously published Luhman 16B map dataset. 
We thank Ben W. P. Lew for providing us with a best-fit exoplanet period for the PPCS-1 in 
Luhman 16B light curves. We thank the anonymous referee for a helpful report.

\bibliographystyle{apj}

\end{document}